# Q Factors Exceeding $10^4$ in Wavelength-to-Subwavelength-Scale Free-Space Resonators

Darrell Omo-Lamai[1,*], Varun Dolia[1], Yanyu Xiong[1], Chih-Yi Chen[1], Parivash Moradifar[1], Priyanuj Bordoloi[1], Sajjad AbdollahRamezani[1], Sahil Dagli[1], Halleh Balch[1], Jennifer Dionne[1,†]

[1]Stanford University, Stanford, CA 94305, USA

*Contact author: deolamai@stanford.edu; †Contact author: jdionne@stanford.edu

**ABSTRACT**. Free-space-addressable optical resonators that combine long photon lifetimes (high $Q$ factors) with strong spatial localization of optical fields (small mode volumes, $V_m$) enhance light–matter interactions with facile far-field excitation. The Purcell factor governing spontaneous emission enhancement scales as $Q\ V_m^{-1}$. Periodically asymmetric resonators, in which perturbations convert bound modes into radiating modes, offer a route to free-space resonances, with the radiative $Q$ factor tuned by the geometric and optical strength of the asymmetry-inducing perturbations. However, free-space resonators that simultaneously achieve high $Q$ and small $V_m$ have remained rare. This limitation arises in part because existing designs do not tailor geometric and optical asymmetries concurrently, thus limiting access to high-$Q$ regimes. Here, we show that jointly tuning geometric and optical asymmetries unlocks a biaxial radiative landscape with iso-$Q$ contours that connect disparate perturbations with equivalent $Q$ factors. We demonstrate this framework with very-large-scale-integrated single-crystalline Si nanoantenna pixels (VINPix) with out-of-plane perturbations of 35–150 nm amorphous Si, $SiN_x$, and $SiO_2$. We experimentally establish biaxial $Q$ factor control in air and achieve $Q$ factors up to 76,000 at wavelength-scale mode volumes ($V_m \sim 1.7\ \lambda_0^3\ n_{eff}^{-3}$) in simultaneously imaged arrays of > 80 resonators in water. Furthermore, we computationally demonstrate 50-nm-wide slotted VINPix that reach $Q$ factors of $10^6$ at subwavelength mode volumes ($V_m \sim 0.2\ \lambda_0^3\ n_{eff}^{-3}$) with 20 nm $SiO_2$ perturbations, yielding Purcell factors as high as $5 \times 10^5$ in an all-dielectric free-space resonator.

Optical resonators with high quality factors ($Q$ factors) and small mode volumes ($V_m$) concentrate electromagnetic energy in space and time, strengthening light–matter interactions for sensing [1,2], nonlinear optics [3], quantum optics [4], and cavity quantum electrodynamics [5]. Together, $Q$ and $V_m$ set the Purcell factor, $F_P$, which quantifies the spontaneous emission enhancement of coupled emitters and scales with $Q\ V_m^{-1}$ [6]. Conventional dielectric resonators, such as photonic-crystal nanocavities [7,8] and whispering-gallery resonators [9], reach large $F_P$ through interference-based confinement and radiation suppression [10,11], but typically require near-field couplers, waveguides, or alignment-sensitive interfaces that preclude wide-field excitation and massively parallel operation.

Free-space resonances are directly accessible from the far field [12]. Such resonances are commonly generated by introducing symmetry-breaking perturbations into periodic high-index dielectric arrays, coupling bound modes to the radiation continuum [13,14]. In the small-perturbation regime, the dimensionless perturbation-induced asymmetry parameter, $\alpha$, tunes the radiative $Q$ factor, with $Q_{rad} \propto \alpha^{-2}$ [15,16]. However, simultaneously achieving high $Q$ and small $V_m$ in free-space resonators has been elusive. $Q$ factors of $10^5$ have been demonstrated, but in spatially extended resonators with large $V_m$ [17,18]. VINPix (very-large-scale-integrated silicon nanoantenna pixels) resonators address the mode volume challenge by truncating a one-dimensionally periodic resonator with photonic crystal mirrors engineered to minimize out-of-plane scattering [19], achieving $V_m \sim 1\ \lambda_0^3\ n_{eff}^{-3}$ in water, which was further reduced to ~ 0.1 $\lambda_0^3\ n_{eff}^{-3}$ with a slot. Yet, experimental $Q$ factors have been limited to 5,000. One reason for this limitation is that the full range of $Q$ accessible through asymmetry scaling has not been systematically explored.

The asymmetry parameter, $\alpha$, contains two physically independent contributions – a geometric asymmetry, $\alpha_g$, set by the spatial magnitude of the perturbation, and an optical asymmetry, $\alpha_o$, set by the refractive index or permittivity magnitude of the perturbation. Despite this two-component structure, most implementations permit the variation of just one component [19–23], requiring geometric and material tolerances at the edge of current fabrication and materials capabilities to minimize $\alpha$ and achieve high $Q$ factors.

Here, we show that independent control of geometric and optical asymmetries unlocks a biaxial landscape of radiative properties in free-space resonators. Mapping $Q$ across this landscape reveals iso-$Q$ contours – families of $(\alpha_g, \alpha_o)$ pairs that realize identical radiative $Q$ factors. Using silicon-compatible free-space-addressable VINPix resonators with wavelength-to-subwavelength mode volumes, we show that biaxial tuning enables broader $Q$ factor tunability than single-parameter designs permit; high-$Q$ resonances rationally designed for different environments, with experimental $Q$ factors up to 76,000 in water; and computed Purcell factors approaching $5 \times 10^5$ in slotted resonators composed solely of standard photonic materials.

We design VINPix resonators for Gaussian-type confinement of an electric dipole mode using photonic crystal mirrors with an explicitly prescribed linear mirror strength progression (see Supplementary Material [25] for methods and details) [7,19,24]. The confined mode is supported by a periodic array of Si nanoblocks (Figure 1(a)). Biperiodic out-of-plane perturbations break inversion symmetry, coupling the mode to the radiation continuum. The height perturbation is given by $\Delta z = z_2 - z_1$, where $z_2$ is the height of the blocks with a perturbation, and $z_1$ is the height of the blocks without a perturbation. This defines the geometric asymmetry, $\alpha_g = \Delta z \times z_1^{-1}$. The refractive index perturbation is given by $\Delta n = n_{per} - n_{env}$, where $n_{per}$ is the refractive index of the perturbation material and $n_{env}$ is the refractive index of the environment. This defines the optical asymmetry, $\alpha_o = \Delta n \times n_{env}^{-1}$. Because the perturbation material is deposited onto the symmetric resonator blocks rather than substituting their composition, $\alpha_g$ and $\alpha_o$ are structurally decoupled variables, and the resonator index $n_{res}$ can be selected independently for mode confinement and spectral positioning.

We realize this biaxial $(\alpha_g, \alpha_o)$ design space on Si-on-sapphire substrates using a two-step electron-beam lithography process in which atomic layer deposition (ALD) controls $\Delta z$ with nanometer precision and the perturbation material library ($SiO_2$, $SiN_x$, $SiN_{x-y}$, amorphous Si) spans $n_{per} \approx 1.45$–$3.3$ (see Supplementary Material [25] for methods). Atomic force microscopy (AFM) topography maps and scanning electron microscopy (SEM) images confirm tunable $\Delta z$ from 25 to 100 nm (Figures 1(b)–(d)). Optical micrographs of resonator arrays with Cr perturbations further illustrate independent $\Delta n$ control (Figure 1(e)).

We first map the biaxial $(\alpha_g, \alpha_o)$ landscape on an infinitely periodic Si-on-sapphire resonator in water ($n_{env} = 1.33$) using full field eigenmode calculations and lossless perturbation materials (Figure 2(a)). $Q$ increases monotonically as the asymmetry is minimized along each axis, diverging as $(\alpha_g, \alpha_o) \rightarrow (0,0)$. For a 5 nm perturbation with $n_{per} = 1.31$ ($\alpha_g = 0.008$, $\alpha_o = 0.015$), $Q = 3.45 \times 10^8$. The product form $\alpha = \alpha_g \times \alpha_o$ collapses this landscape onto the familiar $Q \propto \alpha^{-2}$ scaling when $\alpha_o$ is defined in terms of permittivity (see Supplementary Material [25], Figure S6).

The $(\alpha_g, \alpha_o)$ map highlights that geometric and optical asymmetries can be minimized cooperatively to achieve high $Q$ factors, but further reveals the presence of iso-$Q$ contours – curves of constant $Q$ that span the $(\alpha_g, \alpha_o)$ space. These manifolds connect geometrically and optically distinct perturbations of identical radiative $Q$, as represented in terms of perturbation height and refractive index in Figure 2(b). They establish that any given $Q$ factor is not the property of a unique perturbation, but of a family of perturbations connected by iso-$Q$ equivalence. Hence, a target $Q$ factor can be achieved by trading $\Delta z$ for $\Delta n$, and vice versa. For example, $Q = 10^5$ can be obtained with a shallow perturbation ($\Delta z = 10$ nm) of a high refractive index material ($n_{per} = 3.48$), a thick perturbation ($\Delta z = 100$ nm) of a lower refractive index material ($n_{per} = 1.50$), or a perturbation of intermediate height and refractive index ($\Delta z = 30$ nm and $n_{per} = 1.80$).

Independent geometric and optical asymmetry control expands accessible $Q$ factor tuning ranges beyond those achievable in single-parameter counterparts. We demonstrate this experimentally using 16 μm × 3 μm VINPix resonators fabricated on 300 nm Si-on-sapphire and measured in air using a normal-incidence reflection setup (see Supplementary Material [25] for methods). Figure 3 shows the resulting spectra and $Q$ factors (see Supplementary Material [25], Figures S7–11, for additional spectra). With $\Delta z \approx 100$ nm, decreasing $\Delta n$ by varying the perturbation material from amorphous Si ($n \approx 3.3$) to $SiN_{x-y}$ ($n \approx 2.7$) and to $SiN_x$ ($n \approx 2.0$) increases the mean $Q$ factor from 220 to 540 and to 1,520 across 90 resonators per condition (Figure 3(a)-3(b)). With the perturbation material fixed at $SiN_x$, reducing $\Delta z$ from 150 nm to 50 nm similarly increases the mean $Q$ from 690 to 3,040 across the same number of resonators (Figure 3(c)-3(d)). These results highlight the statistical robustness of biaxial $Q$ factor tuning and the VINPix platform.

The biaxial ($\alpha_g$, $\alpha_o$) framework also provides a rational design strategy for high-$Q$ resonances in a target environment. While $\alpha_g$ can be minimized by decreasing the perturbation height toward zero, $\alpha_o$ depends on the choice of perturbation material relative to the medium. Hence, for any given environment, selecting a material with $n_{per} \rightarrow n_{env}$ independently drives $\alpha_o$ toward zero, resulting in high $Q$ factors. Full field eigenfrequency calculations confirm that $Q$ diverges as $n_{per} \rightarrow n_{env}$ in air, water, and silicon nitride environments (see Supplementary Material [25], Figure S12), validating this strategy in practically relevant media. We use this principle to experimentally demonstrate high-$Q$ VINPix resonators in water. $SiO_2$ (n ≈ 1.45) is a CMOS-compatible dielectric that provides a small optical asymmetry ($\alpha_o$ ≈ 0.06) in this environment. We thus fabricate 15 μm × 2.5 μm VINPix resonators with 35 nm $SiO_2$ perturbations on 600 nm Si-on-sapphire and characterize them in water using a hyperspectral imaging approach (see Supplementary Material [25] for methods). Acquiring spatially resolved frames at a series of wavelengths yields simultaneous spectra for all resonators in the field of view (Figure 4(a)).

Each spectrum is fit to a Fano resonance (Figures 4(b)–(c); see Supplementary Material [25], Figure S13 for additional spectra), resulting in $Q$ factors of 18,300–30,100 across an 82-resonator array, with a mean of 21,300, a standard deviation of 2,400, and a 4 nm spread of resonance wavelengths (Figure 4(d)). Eigenfrequency calculations incorporating fabrication disorder (a normal distribution of block-to-block dimensional variations with $\sigma$ = 3 nm) reduce the ideal $Q_{rad} \approx 6 \times 10^5$ to the $10^4$ range, accounting for the experimentally observed $Q$ factor distribution (see Supplementary Material [25] for details). Individual resonators attain experimental $Q$ factors up to 76,000 (Figure 4(e)), a 15 × improvement over the prior free-space state of the art in water [19]. These high $Q$ factors are achieved on calculated mode volumes of $V_m \approx 1.7\ \lambda_0^3\ n_{eff}^{-3}$ (see Supplementary Material [25], Figure S15). Combining the ($\alpha_g$, $\alpha_o$) framework with approaches to further localize the mode spatially opens a route to free-space resonators with simultaneously high $Q$ factors, subwavelength $V_m$, and large Purcell enhancement factors.

The biaxial framework is directly compatible with slotted resonators that tighten $V_m$. Introducing a 50 nm slot into the VINPix resonator concentrates the electric field of the electric dipole mode in the low-index gap (Figure 5(a)) [19,26–28]. Nevertheless, ($\alpha_g$, $\alpha_o$) tuning continues to govern $Q$ independently. Full field electromagnetic calculations of 16 μm × 3.5 μm slotted VINPix resonators in water show that ($\alpha_g$, $\alpha_o$) tuning spans more than three decades in radiative $Q$ while maintaining $V_m < 0.3\ \lambda_0^3\ n_{eff}^{-3}$ (Figures 5(b)–(c)). With 20 nm $SiO_2$ perturbations, $Q = 10^6$ and $V_m = 0.21\ \lambda_0^3\ n_{eff}^{-3}$, producing a peak Purcell factor $F_P = 4.83 \times 10^5$ for a polarization-matched dipole at the cavity field maximum (Figures 5(d)–(e)). These values approach those of state-of-the-art photonic crystal nanocavities [29] yet arise in a pixelated resonator that preserves dipole-like far-field scattering.

The biaxial asymmetry framework introduced here establishes that the radiative $Q$ factor of asymmetry-driven free-space resonators is governed not by a single parameter, but by a biaxial landscape in which iso-$Q$ contours connect geometrically and optically distinct perturbations that are radiatively equivalent. Using silicon-compatible VINPix resonators, we demonstrate that this framework expands $Q$ factor tunability ranges, enables rational high-$Q$ design in aqueous environments, and supports high Purcell factors when combined with subwavelength mode confinement. The framework is generalizable across resonators in which geometric and optical asymmetries act independently on the same mode. Furthermore, iso-$Q$ contours enable $Q$ to be engineered by trading perturbation geometry against perturbation refractive index, creating opportunities for functional perturbations with application-specific properties, such as electro-optic reconfigurability, emitter compatibility, and surface chemistry. Combining geometric and optical asymmetry control for high $Q$ with spatial mode localization for small $V_m$ broadens the design space for free-space-addressable resonators with strong light–matter interactions, opening practical routes to massively parallel biosensing, nonlinear photonics, and cavity quantum electrodynamics.

We thank Professor Mark Brongersma for insightful discussions that contributed to the theoretical and experimental design. We acknowledge funding from the Moore Foundation, the Office of Naval Research under the Multi-University Research Initiative (MURI) program, and the U.S. Department of Energy Office of Science National Quantum Information Science Research Centers as part of the Q-NEXT center. Part of this work was performed at nano@stanford (RRID:SCR_026695**)**.

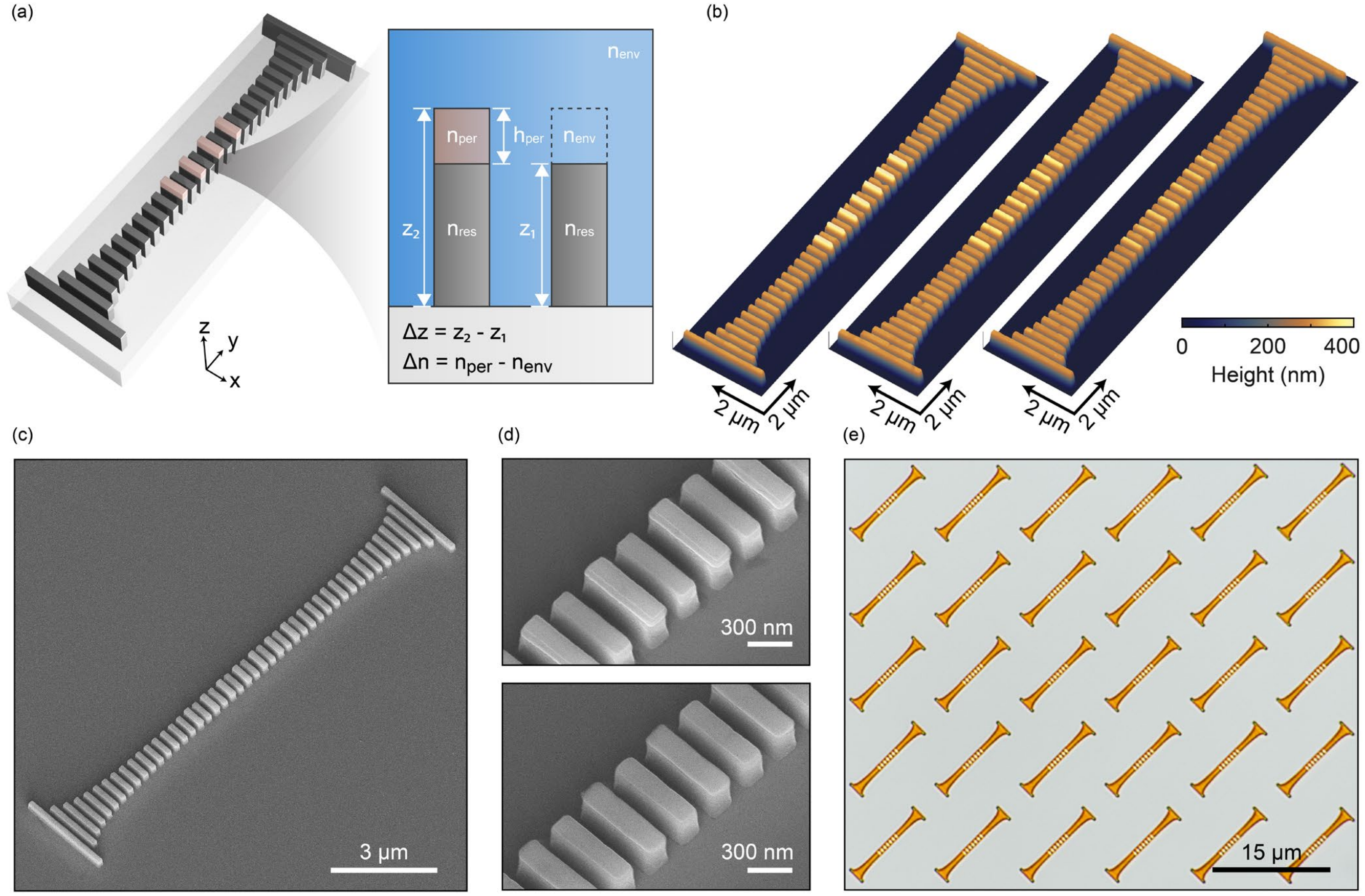

**Figure 1: Independent control of geometric and optical asymmetries on VINPix resonators.** (a) Conceptual framework for geometric and optical asymmetry tuning on VINPix resonators with engineered photonic mirrors. (b) AFM topography maps of resonators with 100 nm (left), 50 nm (middle), and 25 nm (right) height perturbations. (c) SEM images of a resonator with 100 nm height perturbations. (d) High-magnification SEM images of the cavity region for resonators with 100 nm (top) and 25 nm (bottom) height perturbations. (e) Bright-field optical microscopy image of a pixelated resonator array with Cr perturbations.

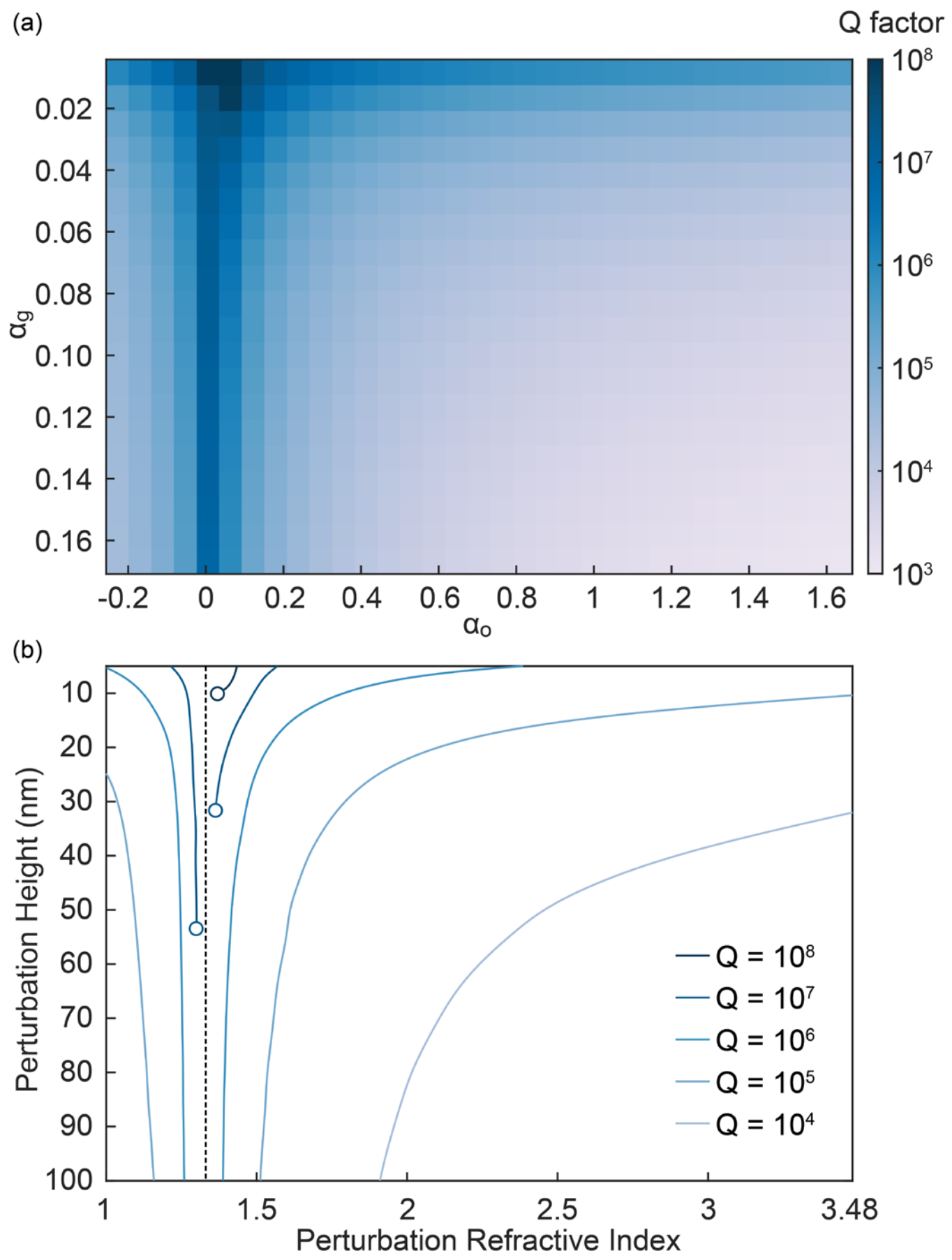

**Figure 2: Simulated biaxial $Q$ factor control via geometric and optical asymmetries.** (a) Map of $Q$ factor as a function of geometric and optical asymmetries ($\alpha_g$ and $\alpha_o$, respectively) for infinitely periodic resonators in water. (b) Iso-$Q$ contours extracted from the ($\alpha_g$, $\alpha_o$) landscape in (a), represented in terms of perturbation height and perturbation refractive index. The plot shows families of perturbations that yield the same $Q$ factors. Discontinuities occur at the perturbation height of 0 nm and the perturbation refractive index of 1.33, as these conditions correspond to the symmetric resonator with an infinite $Q$ factor.

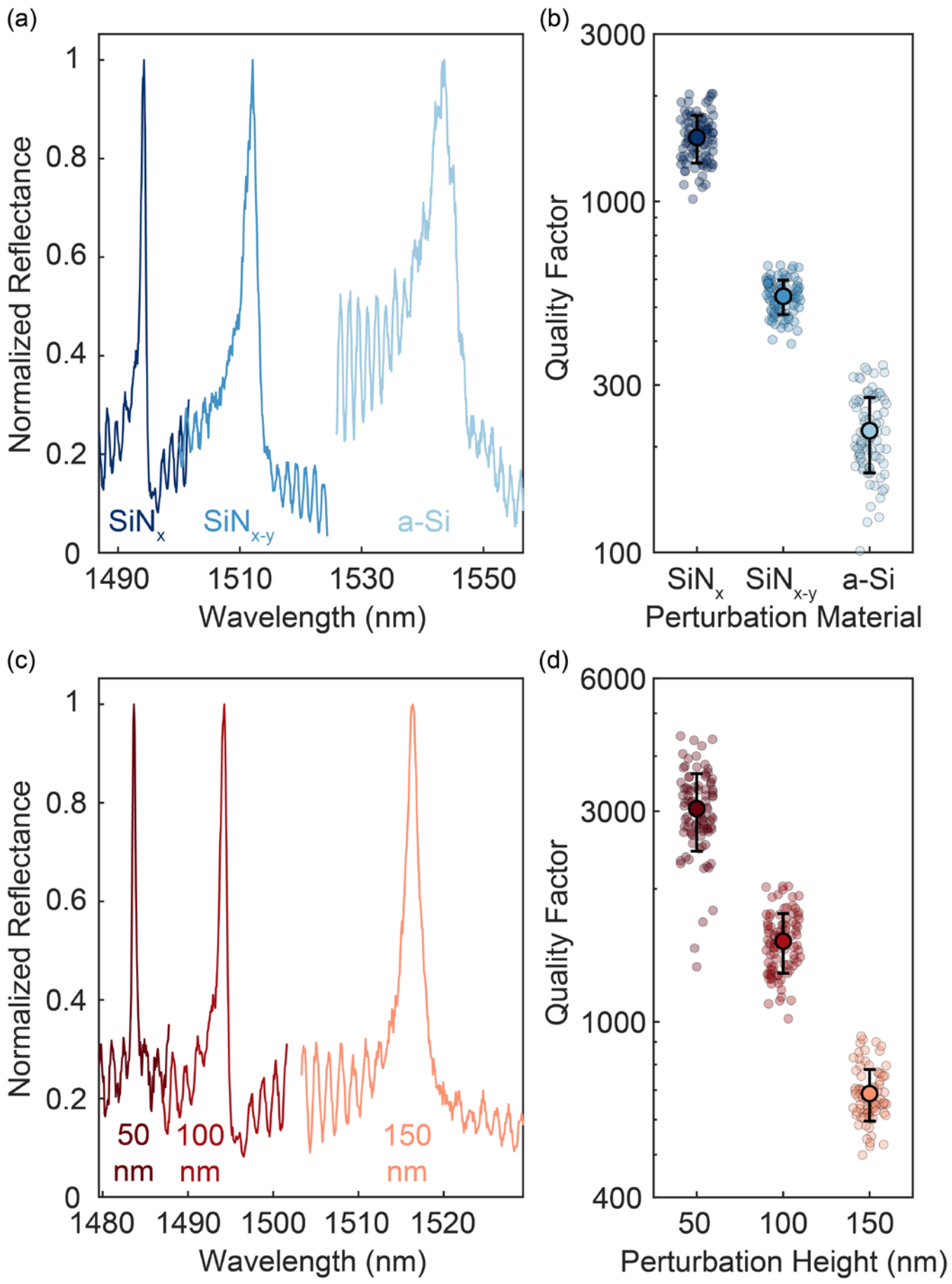

**Figure 3: Experimental $Q$ factor control via perturbation height and material.** (a) Representative reflection spectra for VINPix resonators with 100 nm perturbations of different materials in air. (b) Smaller refractive index asymmetries result in higher average quality factors across 90 resonators per condition. (c) Representative reflection spectra for VINPix resonators with silicon nitride perturbations of different heights in air. (d) Smaller height asymmetries result in higher average quality factors across 90 resonators per condition. Error bars in (b) and (d) correspond to 1 $\sigma$ standard deviations.

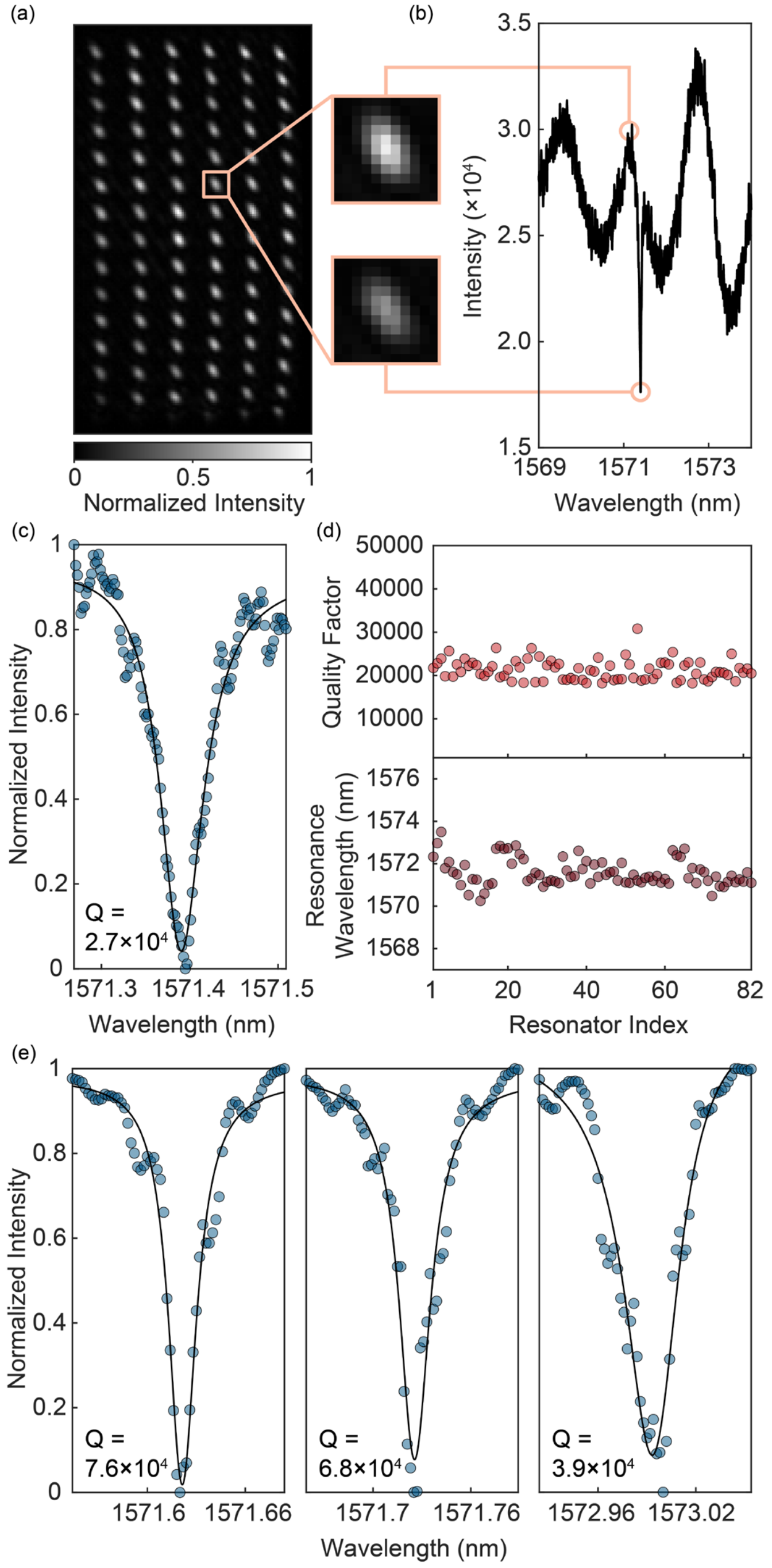

**Figure 4: High-$Q$ VINPix arrays in water.** (a) Wide-field optical micrograph of an array of 84 VINPix resonators with 35 nm $SiO_2$ perturbations, showing an on-resonance intensity decrease for a selected resonator (insets). (b) Extracted spectrum for the selected resonator in (a). Circular markers indicate the wavelengths and intensities corresponding to the micrographs in insets in (a). (c) Fano fit for the resonator in (b), showing a quality factor of 26,900. (d) Summary of quality factors and resonance wavelengths across 82 resonators in the array. (e) Representative spectra for select resonators demonstrating $Q$ factors up to 76,000 (additional examples: $Q$ = 68,000 and 39,000).

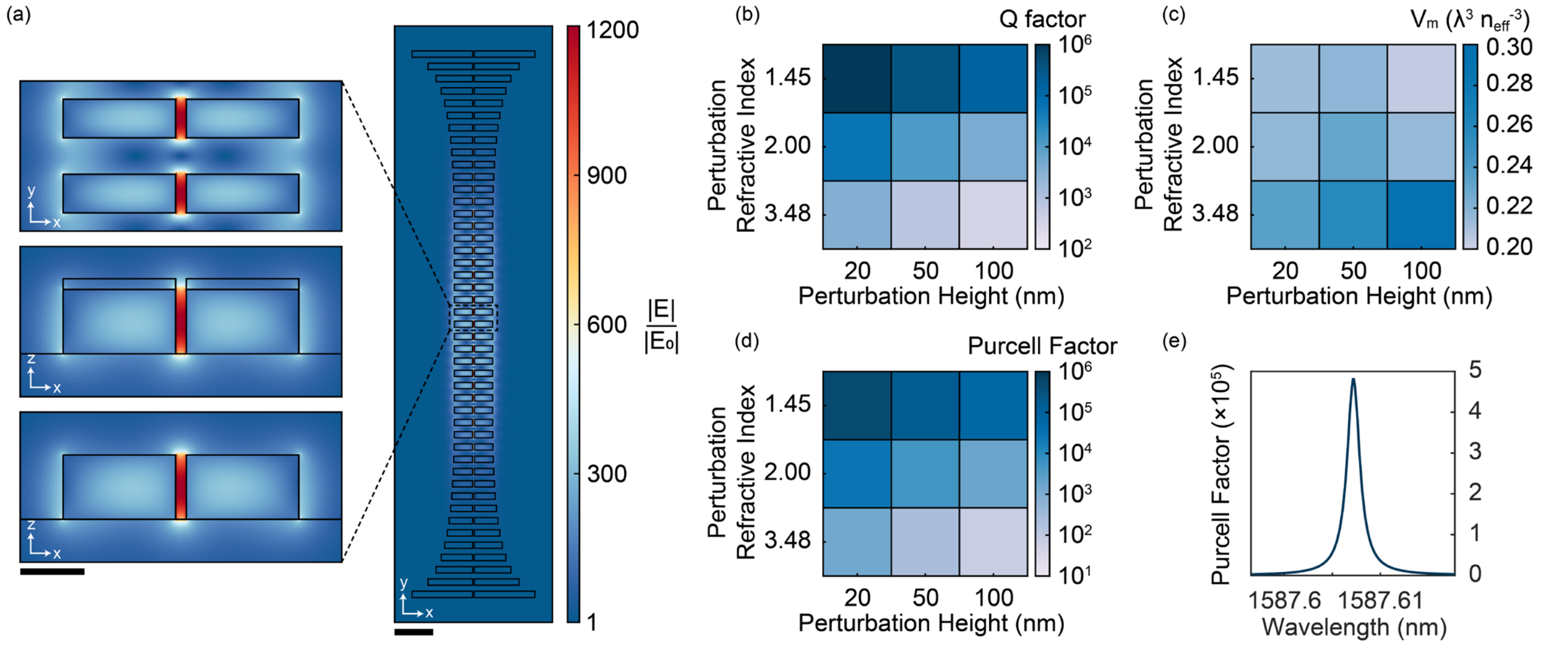

**Figure 5: Slotted VINPix resonators with high $Q$ factors, subwavelength mode volumes, and high Purcell factors.** (a) Simulated electric near-field enhancements at the cross-sections of a 16 µm × 3.5 µm VINPix with 50 nm $SiO_2$ perturbations and a 50-nm-wide slot, showing high enhancements in the low-index gap. Scale bar, 1.1 µm (300 nm in inset). (b) Calculated radiative $Q$ factor as a function of perturbation height and perturbation refractive index for slotted VINPix resonators. (c) Corresponding mode volumes. (d) Corresponding Purcell factors. (e) Simulated Purcell spectrum for a resonator with 20 nm $SiO_2$ perturbations, showing a maximum of $4.83 \times 10^5$ at the resonance wavelength.

# Supplementary Material:
# Q Factors Exceeding $10^4$ in Wavelength-to-Subwavelength-Scale Free-Space Resonators

Darrell Omo-Lamai[1,*], Varun Dolia[1], Yanyu Xiong[1], Chih-Yi Chen[1], Parivash Moradifar[1], Priyanuj Bordoloi[1], Sajjad AbdollahRamezani[1], Sahil Dagli[1], Halleh Balch[1], Jennifer Dionne[1,†]

[1]Stanford University, Stanford, CA 94305, USA

*Contact author: deolamai@stanford.edu; †Contact author: jdionne@stanford.edu

## Supplementary Note I. Methods

### Numerical Calculations

All numerical simulations were performed in COMSOL Multiphysics using the Electromagnetic Waves, Frequency Domain interface. Resonators were modeled on a sapphire substrate, consistent with the experimental platform.

Infinitely periodic resonators were simulated using periodic boundary conditions along the in-plane directions (x and y) and perfectly matched layers (PMLs) along the out-of-plane direction (z). Radiative quality factors were obtained from full field eigenfrequency calculations, where the complex eigenfrequency, $\omega = \omega_0 + i\gamma$, yields $Q = \omega_0\,(2\gamma)^{-1}$. Spectra and near-field distributions were computed using full field wavelength domain studies, with x-polarized plane wave illumination through the sapphire substrate.

Pixelated resonators were simulated using symmetries to reduce computational cost. Perfect electric conductor (PEC) symmetry was applied in x, and perfect magnetic conductor (PMC) symmetry was applied in y (the direction of nanoblock periodicity), with PMLs in z. Exterior boundaries were treated using scattering boundary conditions. Eigenmode analyses were performed using full field eigenfrequency studies. Near-field distributions were computed using scattered field wavelength domain studies.

Mode volumes for pixelated resonators were calculated using the conventional definition, where $V_m$ is determined using the magnitude of the electric field ($E$) and the permittivity ($\varepsilon$):

$$V_m = \int \frac{\varepsilon |E|^2}{\max\,(\varepsilon |E|^2)}$$

The calculated mode volumes were normalized as $V_m\,(\lambda_0^3\, n_{eff}^{-3})$, where $\lambda_0$ is the resonance wavelength of the corresponding mode and $n_{eff}$ is the effective refractive index of the overall structure, determined as the average between resonator, medium, and substrate refractive indices, weighted by the fractional field overlap in each region.

To calculate the Purcell factor, an electric point dipole was positioned at the center of the slot in the pixelated resonator and oriented along x, corresponding to the excitation polarization of the resonator mode. Full field wavelength domain simulations were carried out using the same boundary conditions described above. The emitted power was determined from the outward time-averaged Poynting flux integrated over a closed spherical surface surrounding the dipole. An identical simulation was then performed for the same dipole in the corresponding resonator-free background geometry, which included the sapphire substrate. The Purcell factor was defined as the ratio of the total radiated power in the two cases,

$$F_P = \frac{P_{res}}{P_0}.$$

### Fabrication

A schematic overview of the fabrication process is provided in Figure S1. Devices were fabricated on 300 nm or 600 nm crystalline silicon (Si) on c-plane sapphire substrates (University Wafer), depending on the specific design.

Hydrogen silsesquioxane (HSQ, XR-1541-006, DuPont) was spun on each substrate at 1000 rpm and baked at 90 °C for ~ 5 min. A charge dissipation layer (e-spacer, Showa Denko) was subsequently spun at 2000 rpm and baked at 90 °C for ~ 2 min. The base patterns defining the nominally symmetric resonators were written using 100 kV electron-beam lithography (EBL) (Raith EBPG 5200+). Patterns included alignment markers for the subsequent perturbation definition step. After exposure, samples were developed in a "salty developer" consisting of 95 % $H_2O$, 4 % NaCl, and 1 % NaOH by weight. The Si was then anisotropically etched using inductively coupled plasma reactive ion etching (ICP-RIE, Oxford III-V etcher) with HBr and $Cl_2$ chemistry. The resist was stripped using 2 % HF in water, followed by cleaning in a piranha solution at 120 °C.

Perturbation materials were deposited after the base Si etch. Plasma-enhanced chemical vapor deposition (PlasmaTherm Shuttlelock PECVD) was used to deposit amorphous Si at 200 °C and silicon nitride compositions ($SiN_x$ and $SiN_{x-y}$) at 350 °C. Plasma-enhanced atomic layer deposition (PEALD, Fiji F202, Cambridge Nanotech) was used to deposit < 50 nm $SiO_2$ at 200 °C. This step provides controlled perturbation thicknesses with nanometer-scale precision, particularly for ALD-grown layers.

A secondary lithography step defined the perturbation layers. 950 PMMA (A8 for 600 nm designs and A6 for 300 nm designs) was spun at 2000 rpm and baked at 150 °C for ~ 5 min. A charge dissipation layer (e-spacer, Showa Denko) was spun at 2000 rpm and baked at 150 °C for ~ 2 min. The perturbation blocks were then written by an aligned 100 kV EBL (Raith EBPG 5200+) exposure referenced to the previously patterned markers and developed in methyl isobutyl ketone/isopropanol (1:3). A 30 nm Cr hard mask was deposited by e-beam evaporation (Kurt J. Lesker LAB18), followed by lift-off in acetone for ~ 5 min. The unprotected perturbation material was etched using RIE (OX-RIE Oxford etcher) with $CHF_3$, $CF_4$, and $O_2$ chemistry for $SiO_2$ and SiN-based films, and ICP-RIE (Oxford III-V etcher) with HBr and $Cl_2$ chemistry for amorphous Si. Finally, the Cr mask was removed using a Cr etchant (Transene 1020).

## Materials Characterization

Atomic force microscopy (AFM) was performed using a Park FX-40 system with a Scout 350 probe (NuNano, tip radius: 5 nm). Scanning electron microscopy (SEM) was performed in InLens mode using a Zeiss Gemini 560 instrument, with typical accelerating voltages of 5 to 10 kV. To mitigate charging, samples were typically coated with ~ 5 nm AuPd prior to SEM imaging. Representative SEM images at key fabrication steps are shown in Figure S2. Ellipsometry measurements were performed using a Woollam M2000 Spectroscopic Ellipsometer. Amorphous Si, $SiN_x$, and $SiO_2$ films were fit using Cody-Lorentz models, while $SiN_{x-y}$ films were fit using a Cauchy model. The fitted optical constants are shown in Figure S3.

## Optical Characterization and Analysis

Optical measurements were performed using a home-built near-infrared (NIR) reflection microscope as previously described [1,2]. Samples were illuminated through the sapphire substrate and imaged with a thermoelectrically cooled InGaAs CCD detector (NiRvana, Princeton Instruments).

Samples measured in air were illuminated with a broadband supercontinuum laser (NKT SuperK EXTREME), and spectra were acquired using a spectrometer (Princeton Instruments SPR-2300) with a 600 g $mm^{-1}$ diffraction grating (blaze wavelength 600 nm, Princeton Instruments).

For samples measured in water, spectra were acquired using a tunable laser (Santec TSL-550) in a hyperspectral imaging configuration. The laser wavelength was swept with a dwell time of 100 ms while synchronizing CCD acquisition per wavelength step, producing a time series of wide-field frames. Each frame represents the spatial intensity map at a single wavelength of illumination. The resulting spectromicroscopy data cube was processed by integrating the pixel intensity over a region of interest centered on each resonator to yield the reflection spectrum for each element, enabling simultaneous readout of many resonators within a field of view.

Experimental spectra were fit to a Fano model that accounts for both the resonant scattering response and Fabry–Pérot interference in the substrate.

$$T = \left|\frac{1}{1 + F\sin^2(n_s k h_s)}\right| \left|a_r + i a_i + \frac{b}{f - f_0 + i\gamma}\right|^2$$

$T$ is the scattered intensity. The prefactor captures Fabry–Pérot modulation through a substrate of thickness $h_s$ and refractive index $n_s$. $k = 2\pi\,\lambda^{-1}$ is the free-space wavevector magnitude and $F$ accounts for the reflectivity of the interfaces. The second term models resonant scattering, where $a_r + i a_i$ represents a constant complex background and the resonant contribution is modeled as a Lorentzian oscillator with resonance frequency $f_0$ and full width at half maximum $2\gamma$. $Q$ factors were extracted as $Q = f_0\,(2\gamma)^{-1}$ from the fitted spectra.

## Supplementary Note II. Mode Confinement Using Photonic Crystal Mirrors

To confine resonances in nominally infinite resonators to a finite cavity while minimizing $Q$ factor degradation due to out-of-plane scattering, we pixelate the resonators using photonic crystal mirrors. It is well established that optimal confinement in such nanocavities is achieved when the cavity mode approaches a Gaussian field envelope, which minimizes the spatial Fourier components that fall within the light cone [3]. In our initial pixelated resonator demonstration, we approximated this condition using a polynomial taper in the mirror block width [1]. Here, we instead design the taper from an explicit relationship between Bragg mirror field attenuation and the target Gaussian envelope [4].

Within a Bragg mirror, the electric field amplitude can generally be written in terms of an oscillatory Bloch component and an exponential decay,

$$E(x) \propto \sin(\beta x)\, e^{-\kappa x},$$

where $\beta$ is the propagation constant and $\kappa$ is the attenuation constant, or mirror strength. A Gaussian cavity envelope may be written in the same form but with a quadratic exponent,

$$E(x) \propto \sin(\beta x)\, e^{-\sigma x^2}.$$

Matching the decay factors implies $\kappa x = \sigma x^2$. Hence,

$$\kappa(x) = \sigma x,$$

so the mirror strength should increase linearly with position away from the cavity. We therefore design the mirror taper by selecting mirror block widths that implement a linear $\kappa(x)$ profile.

For each candidate mirror block width, $w$, we compute the relevant band frequencies for the mode of interest in a corresponding periodic mirror section via eigenmode analysis. These are the lower and upper band edge frequencies, $\omega_1(w)$ and $\omega_2(w)$, the mid-gap frequency, $\omega_0(w)$, and the resonance frequency, $\omega_{res}$, set by the cavity region. The mirror strength is defined as

$$\kappa(w) = \sqrt{\left(\frac{\omega_2 - \omega_1}{\omega_2 + \omega_1}\right)^2 - \left(\frac{\omega_{res} - \omega_0}{\omega_0}\right)^2}$$

Using this approach, we design photonic crystal mirrors to confine an electric dipole mode supported by a periodic 600 nm Si-on-sapphire resonator (Figure S4). Figure S5(a) shows $\omega_1(w)$, $\omega_2(w)$, $\omega_0(w)$, and $\omega_{res}$ as functions of mirror block width for the resonator, and Figure S5(b) gives the corresponding $\kappa(w)$.

We choose a maximum mirror block width $w_{max}$ = 2.5 μm, beyond which $\kappa$ increases only weakly (Figure S5(b)), and set the minimum mirror block width $w_{min}$ equal to the nominal cavity block width to ensure a continuous transition at the cavity-mirror interface, $w_{min}$ = 0.6 μm. To construct the taper, we prescribe a set of linearly spaced mirror strengths between $\kappa(w_{min})$ and $\kappa(w_{max})$, and then select the corresponding widths by inverting the numerically obtained $\kappa(w)$ curve (as illustrated by the lines in Figure S5(b)). This procedure yields a taper with an explicitly linear progression of mirror strengths, consistent with the Gaussian envelope condition.

To determine the number of mirror blocks required for practical convergence to the infinite resonator radiative limit, we simulate a representative pixelated resonator design with a 7-period cavity consisting of 10 nm $SiO_2$ height perturbations while varying the number of mirror blocks on each side. Figure S5(c) shows that the $Q$ factor increases with mirror length and approaches the value in the equivalent infinite resonator. Based on this convergence behavior, we use 15 mirror blocks per side, with the linear $\kappa$ progression described above. The finalized device comprises a 4.95 μm cavity with 4.95 μm tapers on each side (Figure S5(d)).

Figure S5(e) shows the simulated electric near-field distribution for the finalized pixelated resonator with 50 nm $SiO_2$ height perturbations in the cavity region, confirming strong spatial localization with peak field enhancements of ~ 1250.

We compare the linear $\kappa$ taper to the previously used fourth-order polynomial taper. Figure S5(f) summarizes the resulting $Q$ factors as a function of perturbation height, $\Delta z$. The linear $\kappa$ design generally yields higher $Q$ factors, with the most pronounced improvements occurring for small perturbations, where mirror leakage and scattering sensitivity are most stringent. This comparison supports the use of an explicitly prescribed linear mirror strength profile as a theoretically robust basis for high-$Q$ confinement in periodic resonators. Using the same approach, we design pixelated 300 nm Si-on-sapphire resonators with a 5.25 μm cavity and 5.25 μm tapers on each side, using $w_{max}$ = 3 μm and $w_{min}$ = 0.7 μm for the standard design, and $w_{max}$ = 3.5 μm, $w_{min}$ = 1.1 μm, and a 50-nm-wide slot for the slotted design.

## Supplementary Note III. Impact of Fabrication Defects on the Radiative Quality Factor

The measured $Q$ factors in Figure 4 are below the radiative limits predicted for ideal structures because experimental devices possess additional loss channels, including absorption in water at L-band wavelengths, surface/interface roughness, and disorder-induced scattering. Here, we isolate geometry-driven reductions in the radiative $Q$ factor by simulating two common fabrication irregularities (schematically shown in Figure S14(a)) – rounded edges and block-to-block dimensional mismatch due to the finite resolution of lithographic patterning.

All calculations are eigenfrequency studies of a representative ~ 15 μm pixelated resonator (4.95 μm cavity, 4.95 μm taper sections) with 35 nm $SiO_2$ height perturbations in the cavity region, matching the devices in Figure 4. Materials remain treated as lossless.

We model systematic rounding by applying a uniform edge radius, $r_{edge}$, to all blocks. As shown in Figure S14(b), the ideal sharp-edged structure ($r_{edge} = 0$ nm) yields $Q \sim 6.1 \times 10^5$. Increasing the rounding to $r_{edge} = 20$, 30, and 50 nm produces only a modest change, with $Q$ remaining $\sim 5.5 \times 10^5$ for $r_{edge} = 50$ nm.

To capture stochastic size variations, we set $r_{edge} =$ 50 nm and perturb each block's areal dimensions by independent random size offsets drawn from a normal distribution with a mean of 0 nm and a standard deviation of 3 nm. We compute the $Q$ factor for 20 disorder realizations (different random seeds). Figure S15(c) shows that this disorder produces a substantially larger suppression of radiative $Q$. Nevertheless, $Q$ factors approaching $10^5$ remain attainable. This is consistent with our experimental results shown in Figure 4(e).

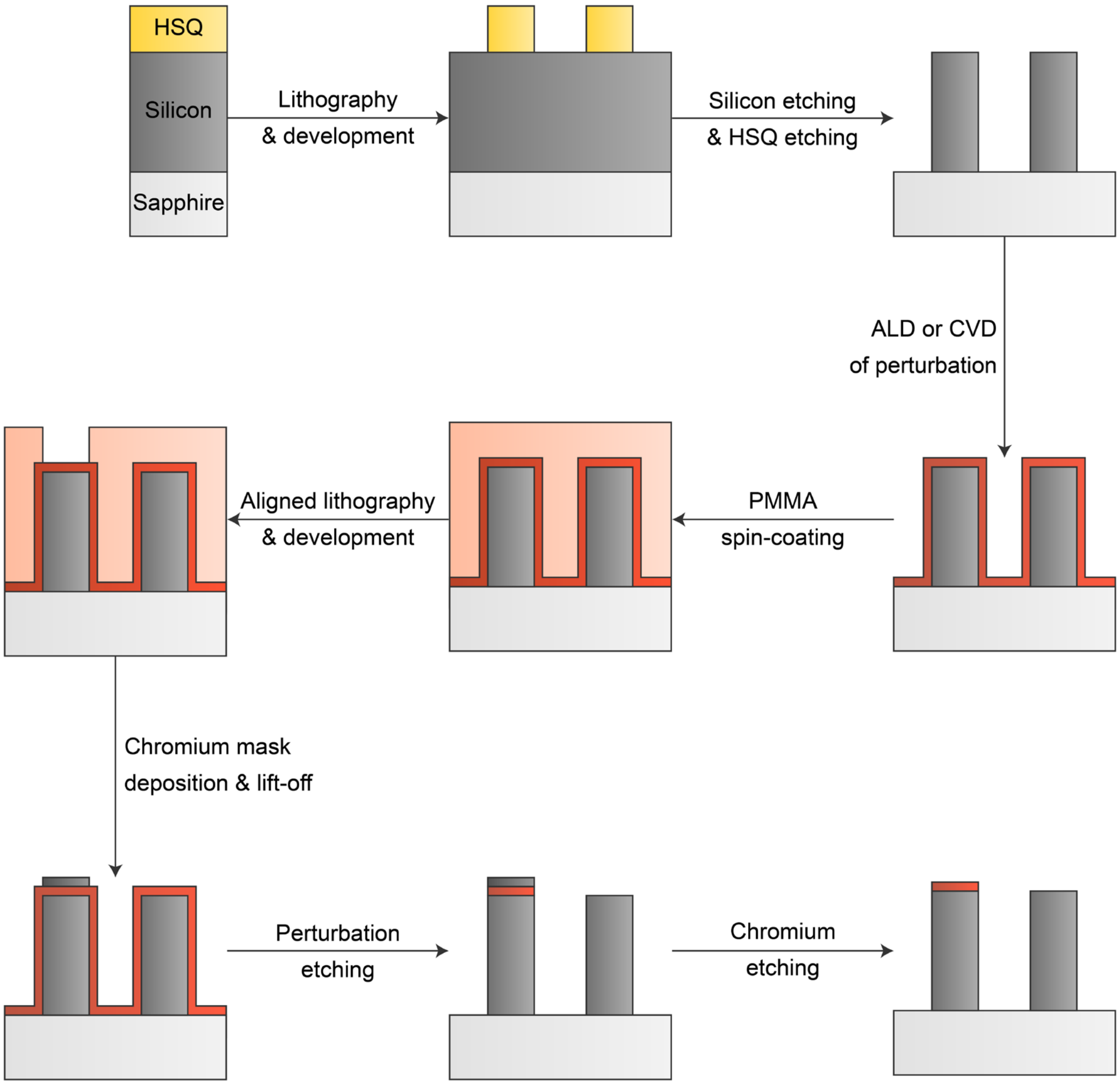

**Figure S1:** Schematic of the multi-step fabrication process for asymmetric resonators with independently tunable perturbation heights and perturbation materials.

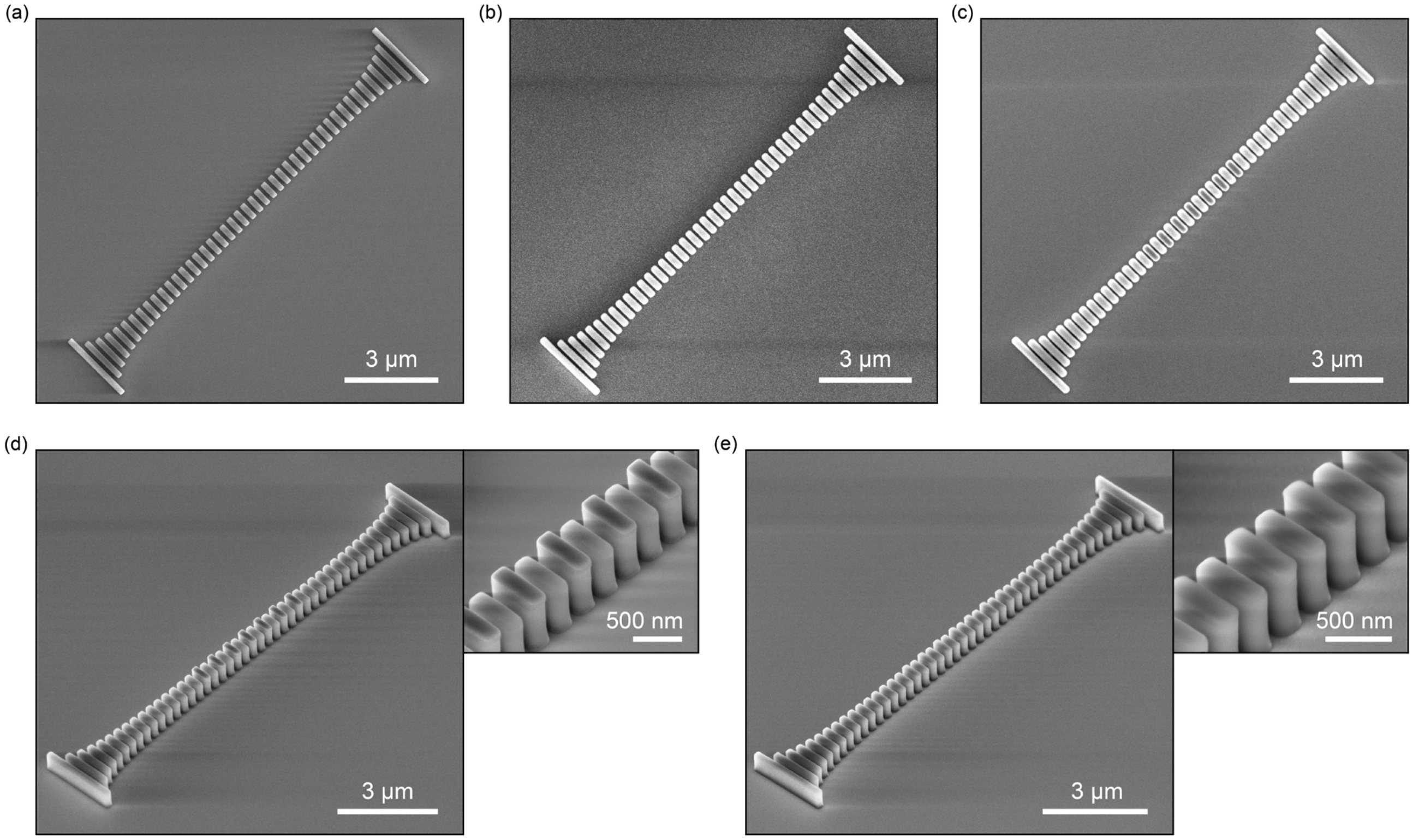

**Figure S2:** Representative SEM micrographs of pixelated resonators after (a) Si etching and HSQ resist removal, (b) perturbation material deposition, (c) Cr hard mask deposition and lift-off, (d) perturbation material etching, and (e) Cr hard mask etching to produce the finalized structure.

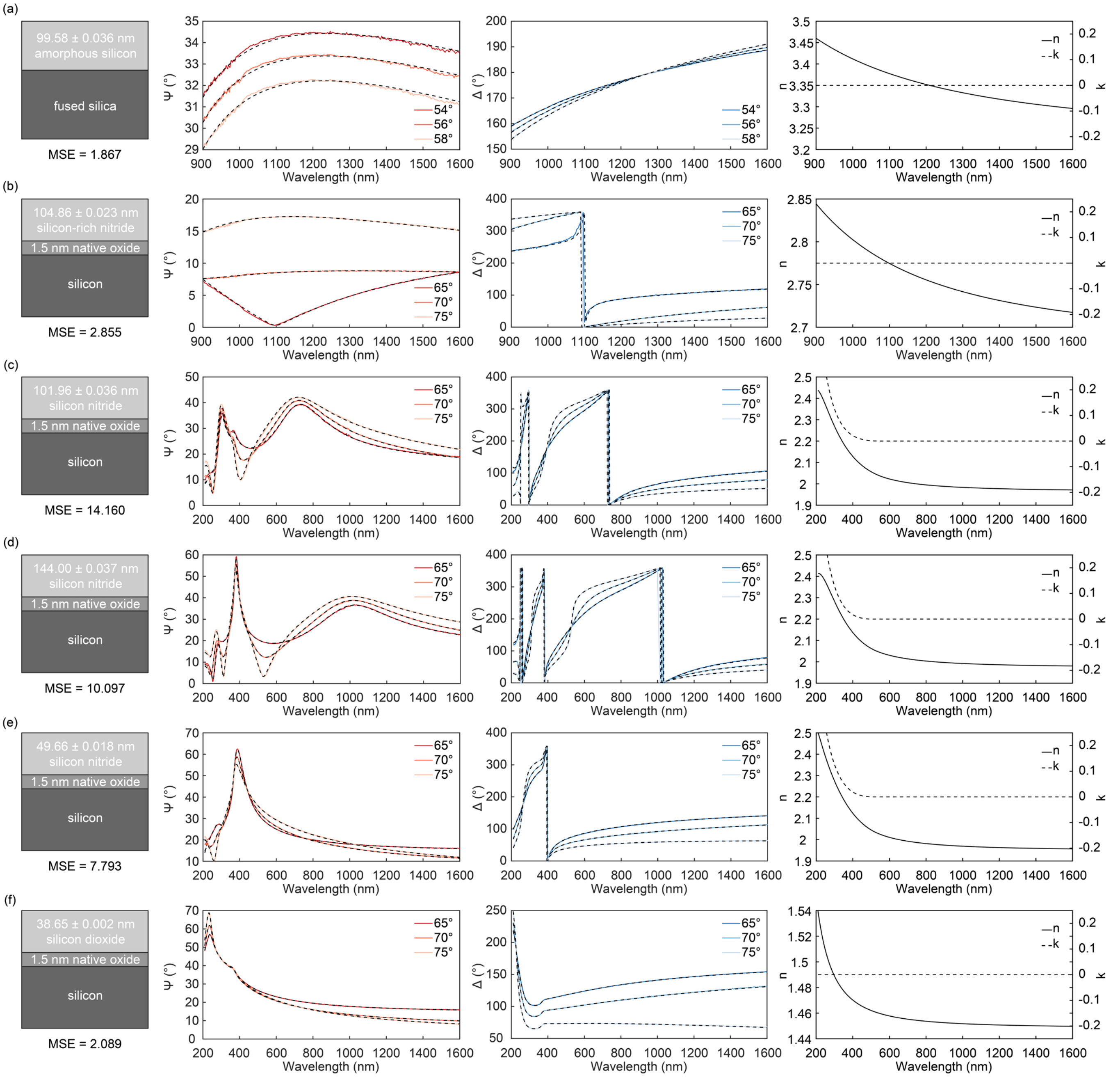

**Figure S3.** Ellipsometry fits for different perturbation materials and heights. Fits were performed on planar chips deposited alongside the resonator chips during fabrication. For each sample, the modeled layer stack, measured (solid) and best fit (dashed) $\Psi$ and $\Delta$ spectra, and extracted optical constants ($n$ and $k$) over the fitting wavelengths are shown.

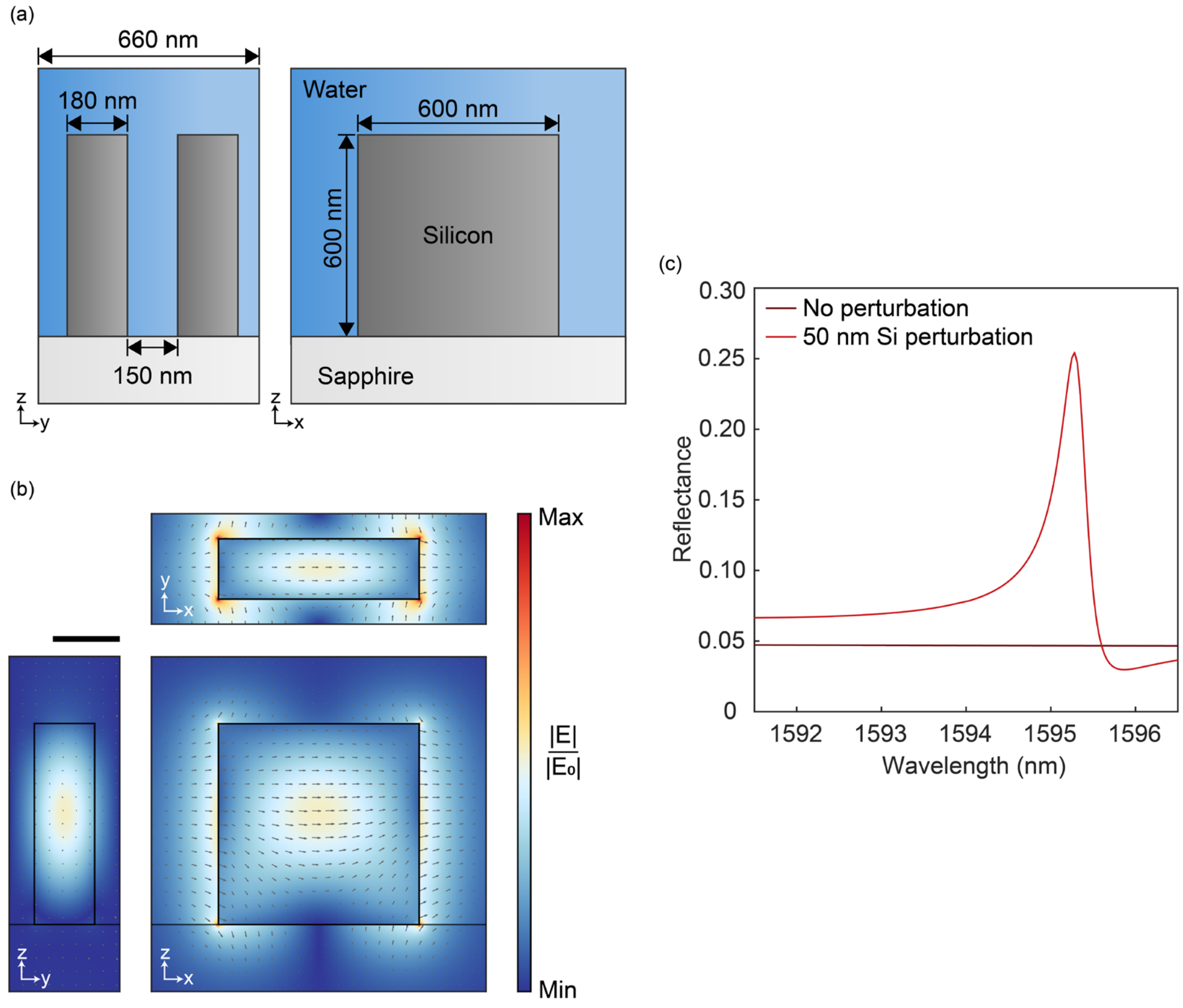

**Figure S4:** (a) Schematic of the unit cell of the 600-nm-tall periodic resonator, with key dimensions indicated. (b) Simulated electric-field intensity distribution of the targeted electric dipole mode in the symmetric resonator. (c) Simulated normal incidence reflection spectrum showing the emergence of a spectral resonance feature upon introducing a biperiodic height perturbation of 50 nm Si.

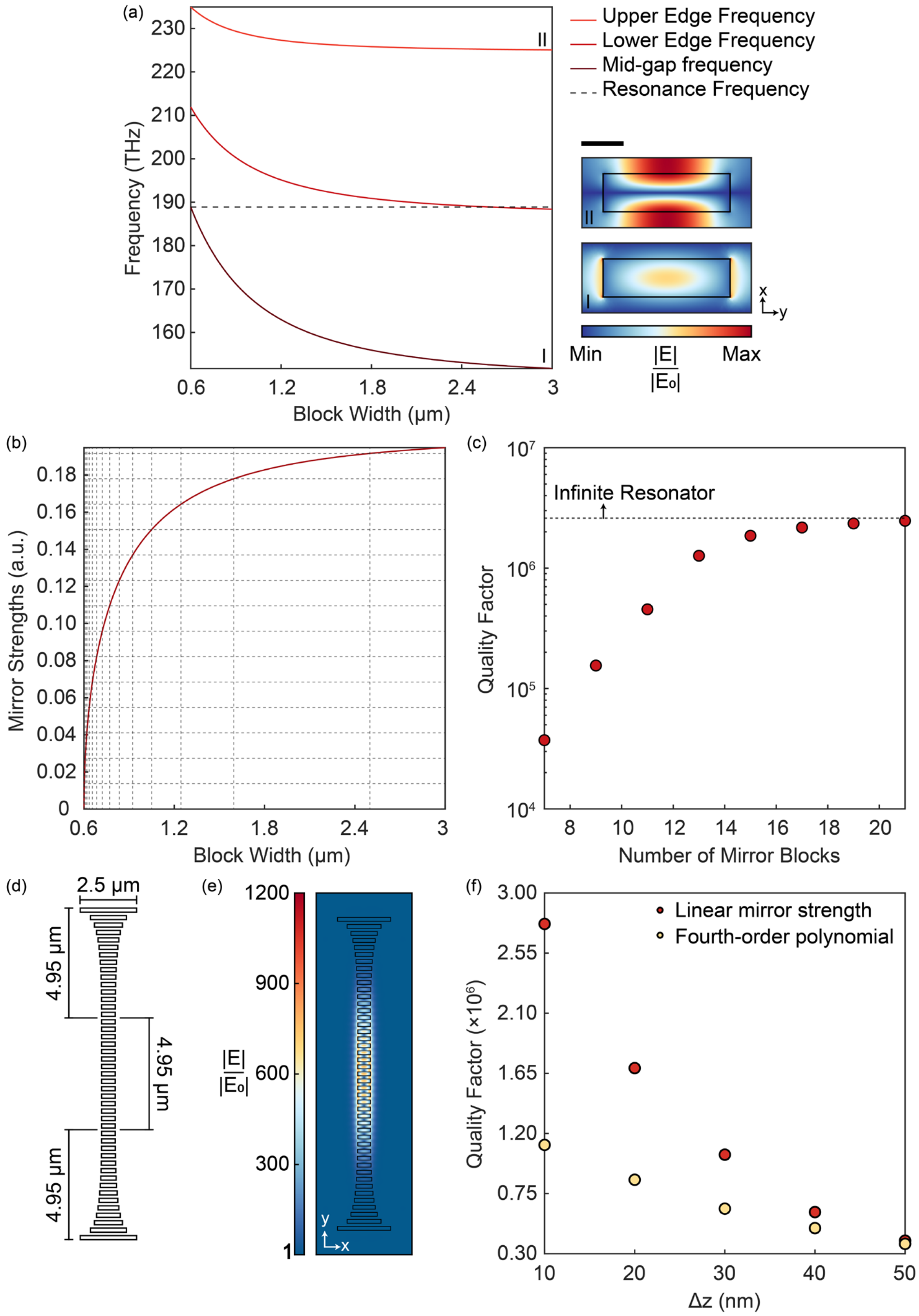

**Figure S5: Design of pixelated resonators with photonic crystal mirrors.** (a) Numerically computed upper and lower band edge frequencies and mid-gap frequencies of the relevant electric dipole mode for different mirror block widths. Representative electric near-field profiles for the upper and lower band edges are shown. (b) Calculated mirror strengths for different mirror block widths. (c) Simulated $Q$ factors for various numbers of terminating mirror blocks (per side) for a resonator with biperiodic 10 nm $SiO_2$ height perturbations in a 5 µm (7-period) cavity, showing convergence toward the infinite resonator radiative limit. (d) Geometry of the finalized 600-nm-tall resonator. (e) Simulated cross-sectional electric near-field enhancements for the finalized 600-nm-tall resonator with 50 nm $SiO_2$ height perturbations, demonstrating strong cavity localization. (f) Comparison of $Q$ factors for pixelated resonators designed by a linear mirror strength progression versus a fourth-order polynomial progression for different $SiO_2$ perturbation heights, showing improved confinement for the linear mirror strength design, particularly at small perturbation magnitudes.

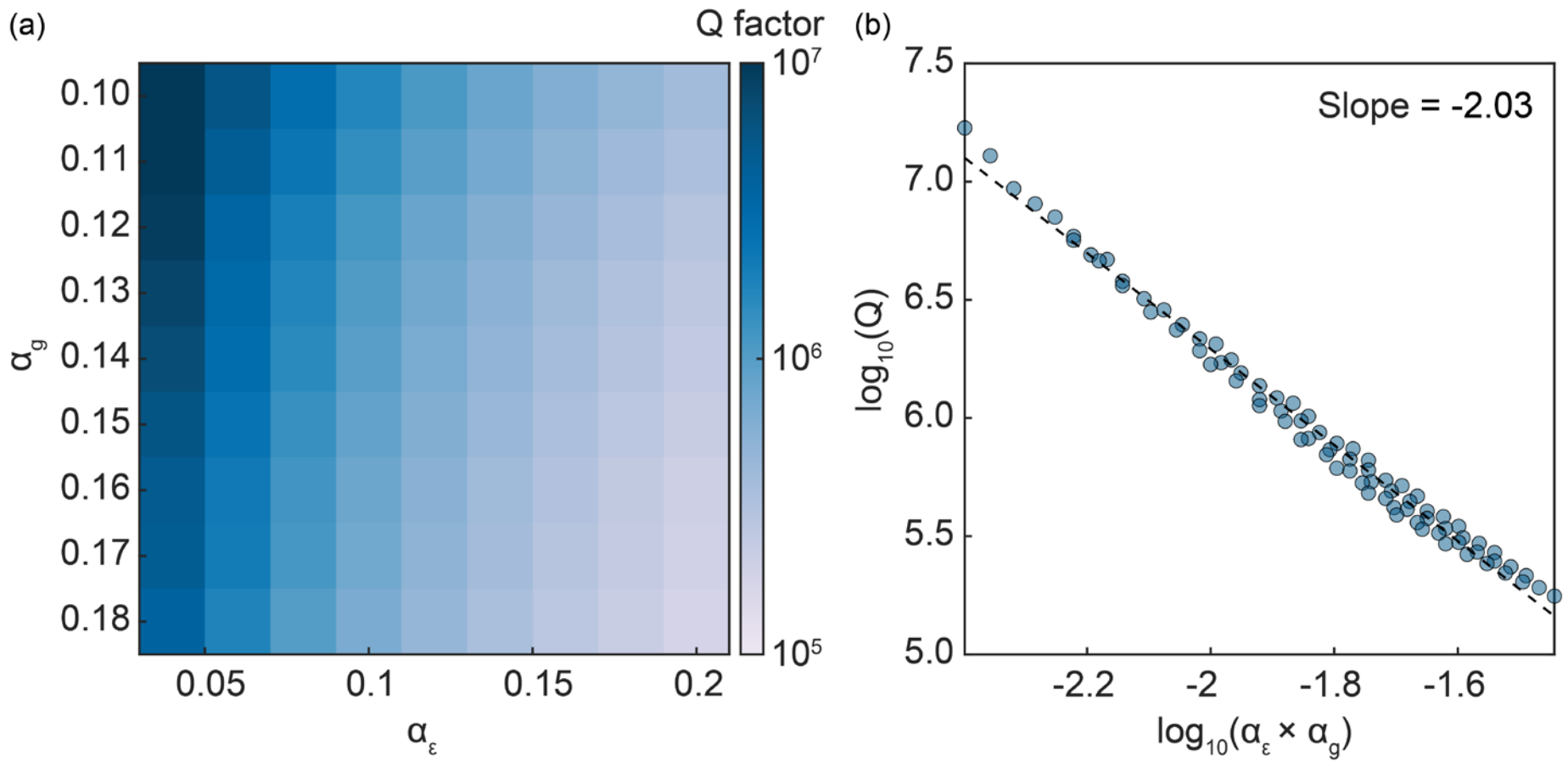

**Figure S6:** (a) Calculated map of $Q$ factor for different geometric asymmetries ($\alpha_g$, defined as $\Delta z \times z_1^{-1}$) and optical asymmetries ($\alpha_\varepsilon$, defined in terms of permittivity as $\Delta\varepsilon \times \varepsilon_1^{-1}$). This map is a subset of that shown in Figure 2(a). (b) Log-log scaling of $Q$ with the generalized asymmetry $\alpha = \alpha_g \times \alpha_\varepsilon$, showing an approximately inverse-square dependence as indicated by the linear line with slope ~ -2.

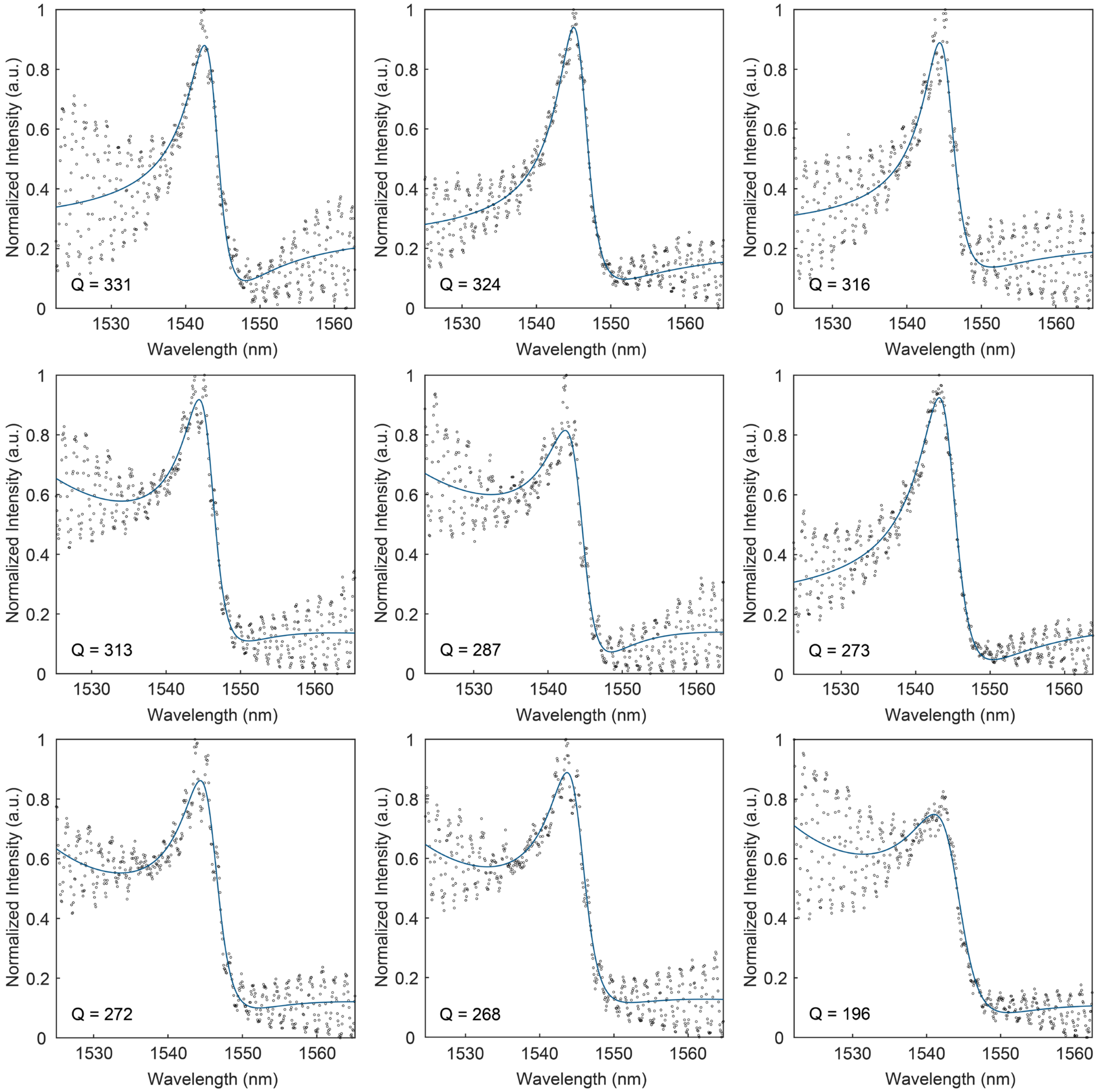

**Figure S7:** Additional representative experimental reflection spectra (markers) and Fano fits (solid lines) for pixelated resonators with ~ 100 nm amorphous Si perturbations measured in air. Extracted $Q$ factors are indicated on each panel.

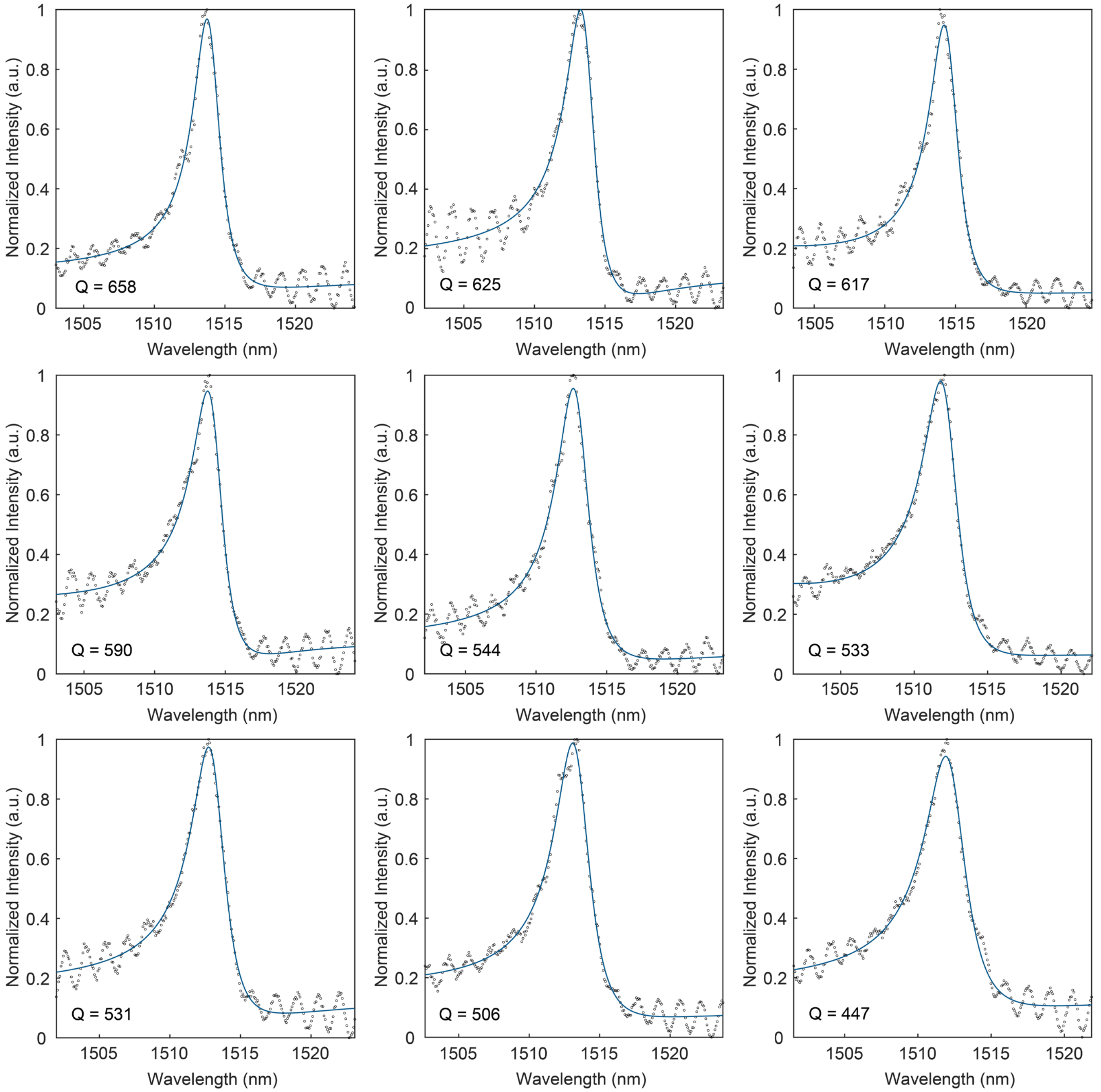

**Figure S8:** Additional representative experimental reflection spectra (markers) and Fano fits (solid lines) for pixelated resonators with ~ 100 nm silicon-rich nitride ($SiN_{x-y}$) perturbations measured in air. Extracted $Q$ factors are indicated on each panel.

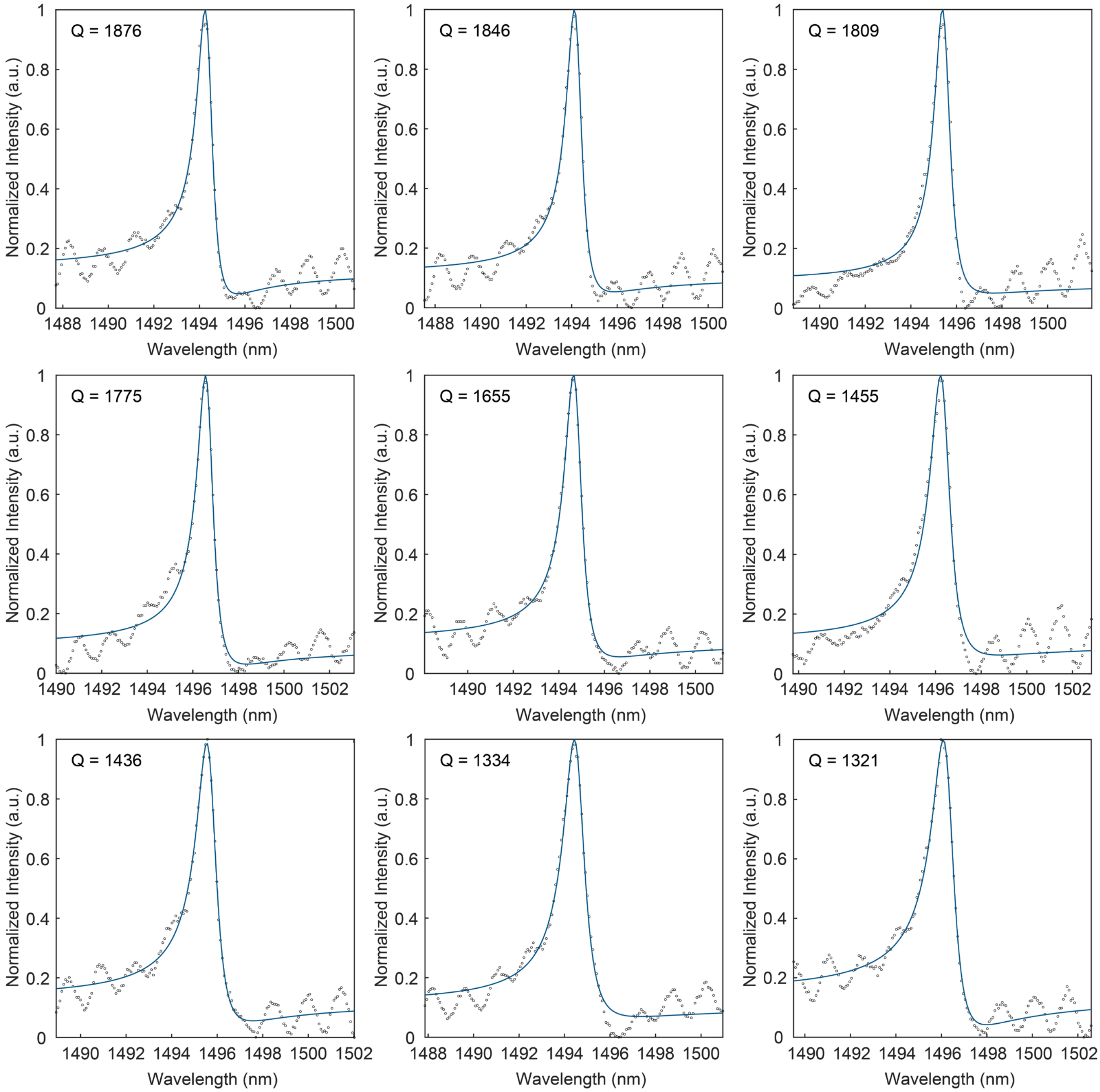

**Figure S9:** Additional representative experimental reflection spectra (markers) and Fano fits (solid lines) for pixelated resonators with ~ 100 nm silicon nitride ($SiN_x$) perturbations measured in air. Extracted $Q$ factors are indicated on each panel.

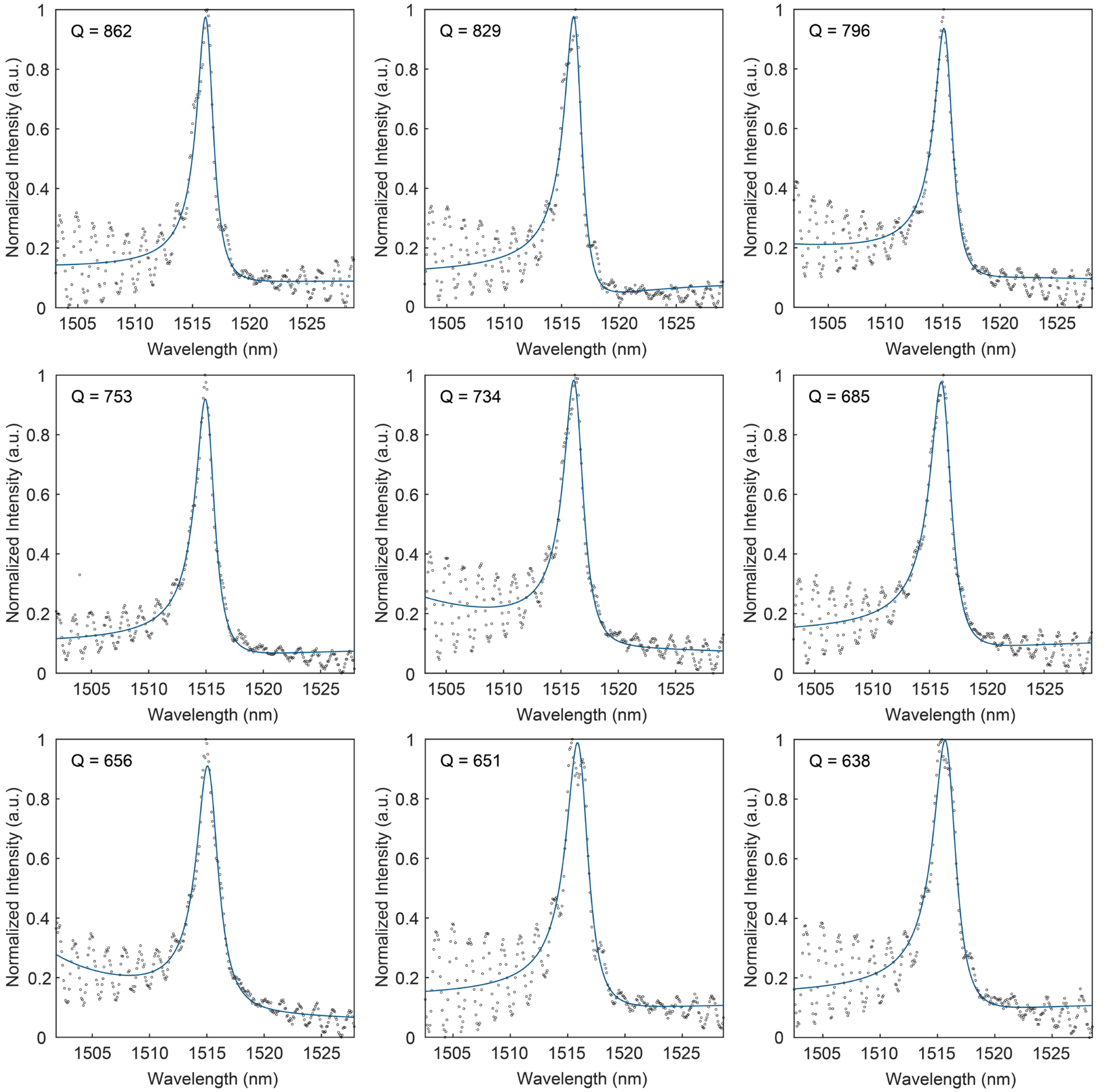

**Figure S10:** Additional representative experimental reflection spectra (markers) and Fano fits (solid lines) for pixelated resonators with ~ 150 nm silicon nitride ($SiN_x$) perturbations measured in air. Extracted $Q$ factors are indicated on each panel.

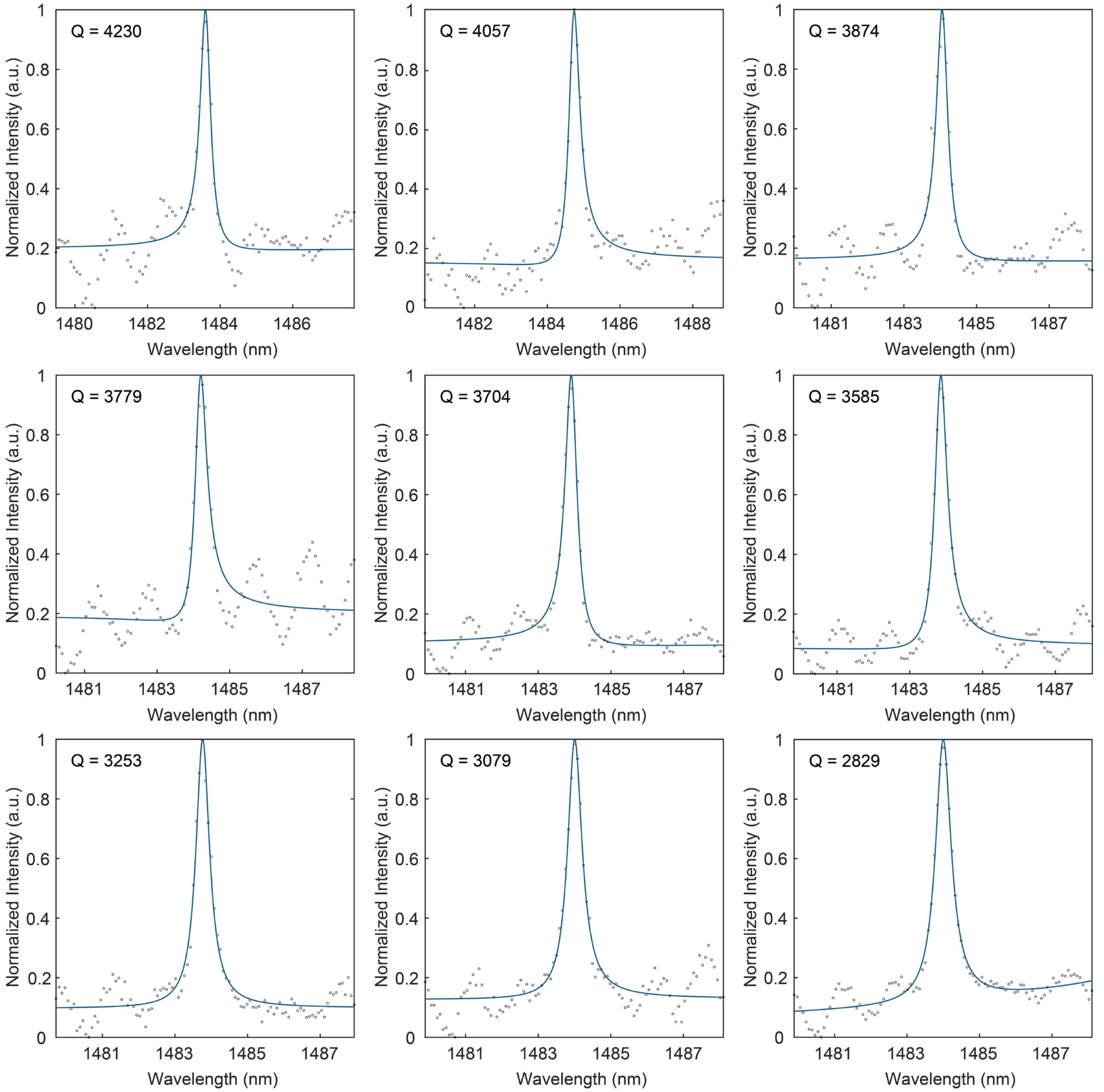

**Figure S11:** Additional representative experimental reflection spectra (markers) and Fano fits (solid lines) for pixelated resonators with ~ 50 nm silicon nitride ($SiN_x$) perturbations measured in air. Extracted $Q$ factors are indicated on each panel.

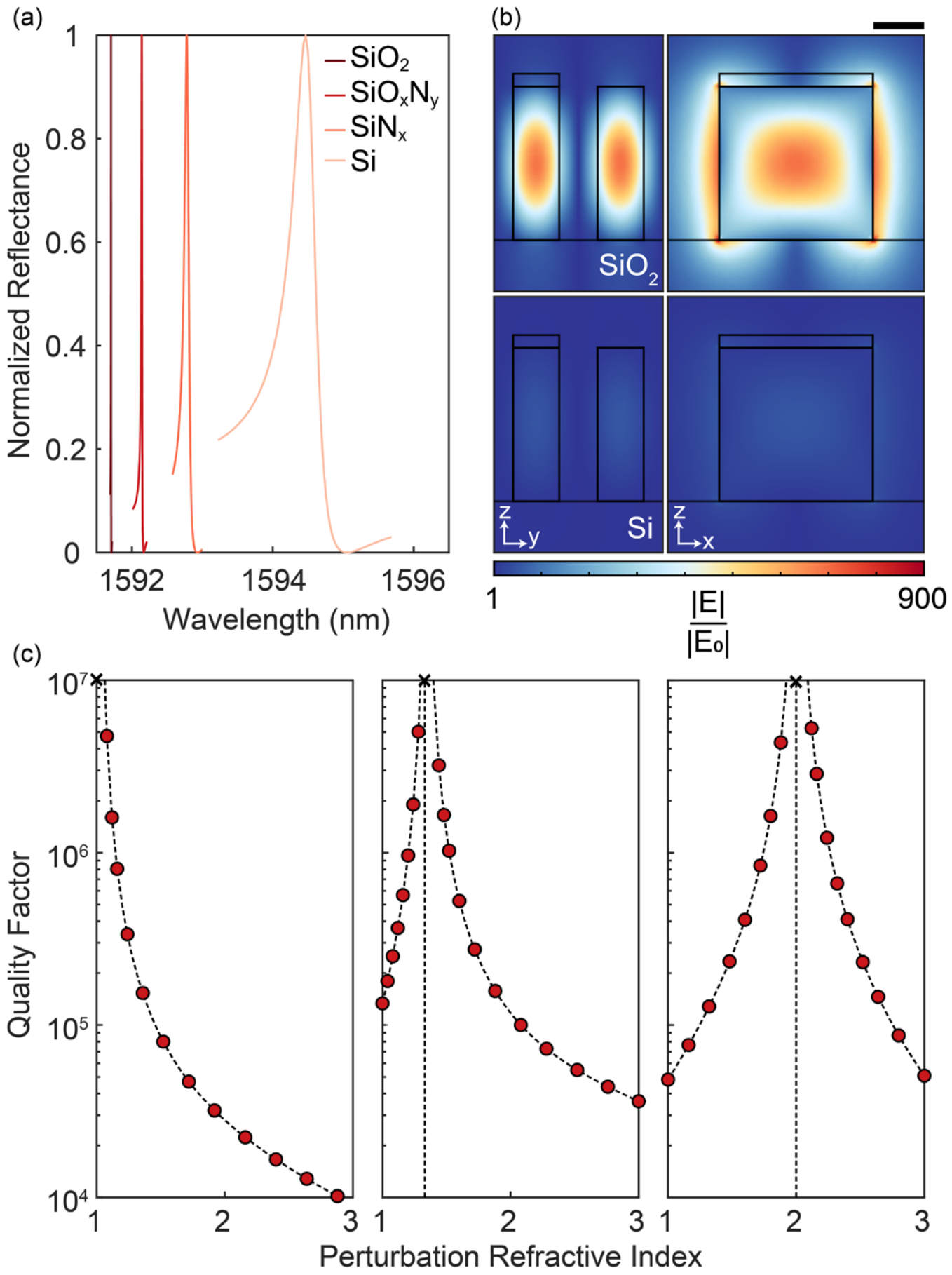

**Figure S12: Simulated high-*Q* resonances across operating environments.** (a) Reflectance spectra for infinite resonators with 50 nm height perturbations of different materials in water. (b) Resonators with $SiO_2$ perturbations (top) have higher cross-sectional electric near-field enhancements than those with Si perturbations (bottom). Scale bar, 200 nm. (c) $Q$ factors for resonators with 50 nm height perturbations of different materials in air (left), water (middle), and silicon nitride (right). The vertical lines indicate the refractive indices of the respective environments.

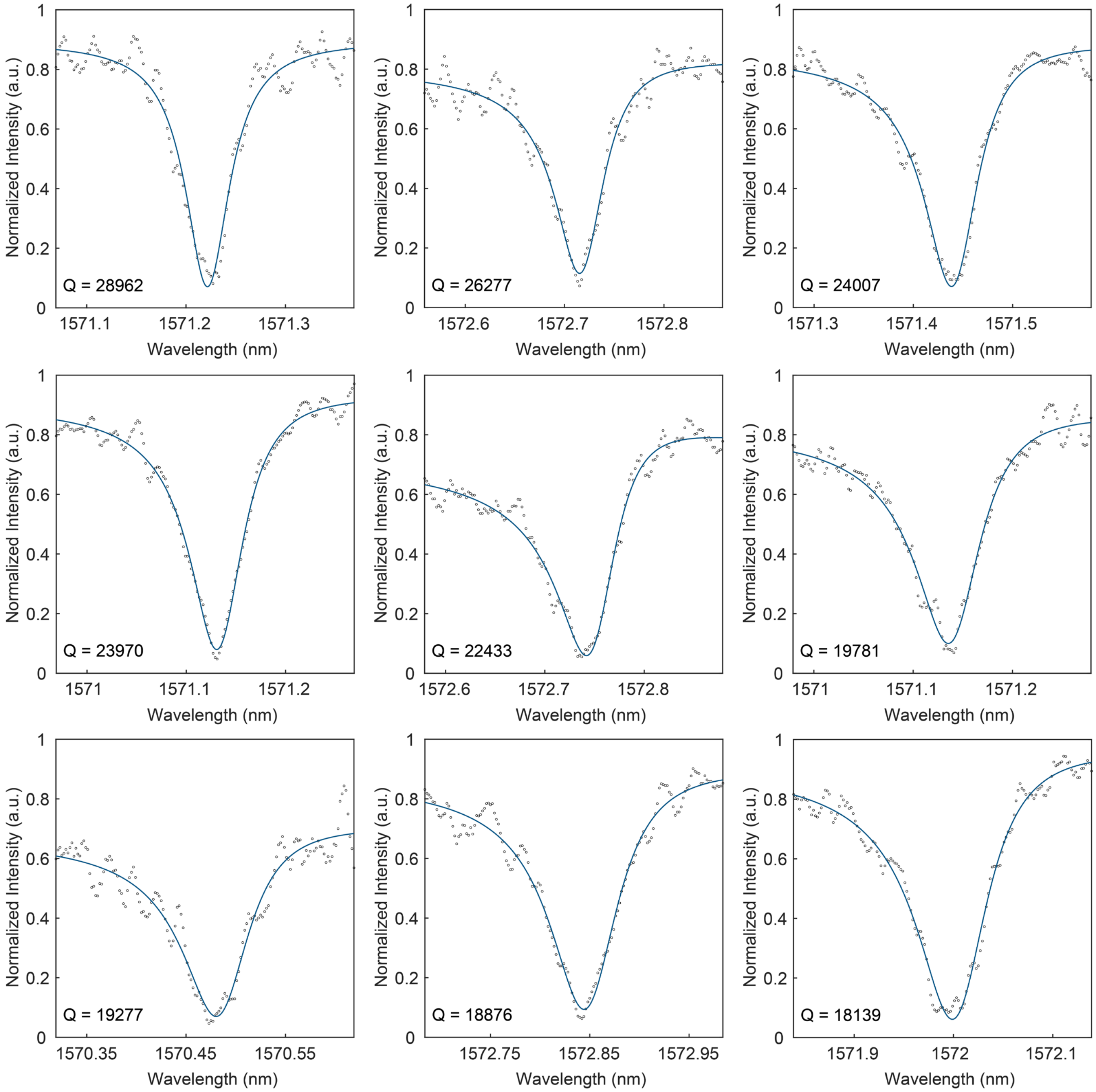

**Figure S13:** Additional representative experimental reflection spectra (markers) and Fano fits (solid lines) for pixelated resonators with ~ 35 nm silicon dioxide ($SiO_2$) perturbations measured in water. Extracted $Q$ factors are indicated on each panel.

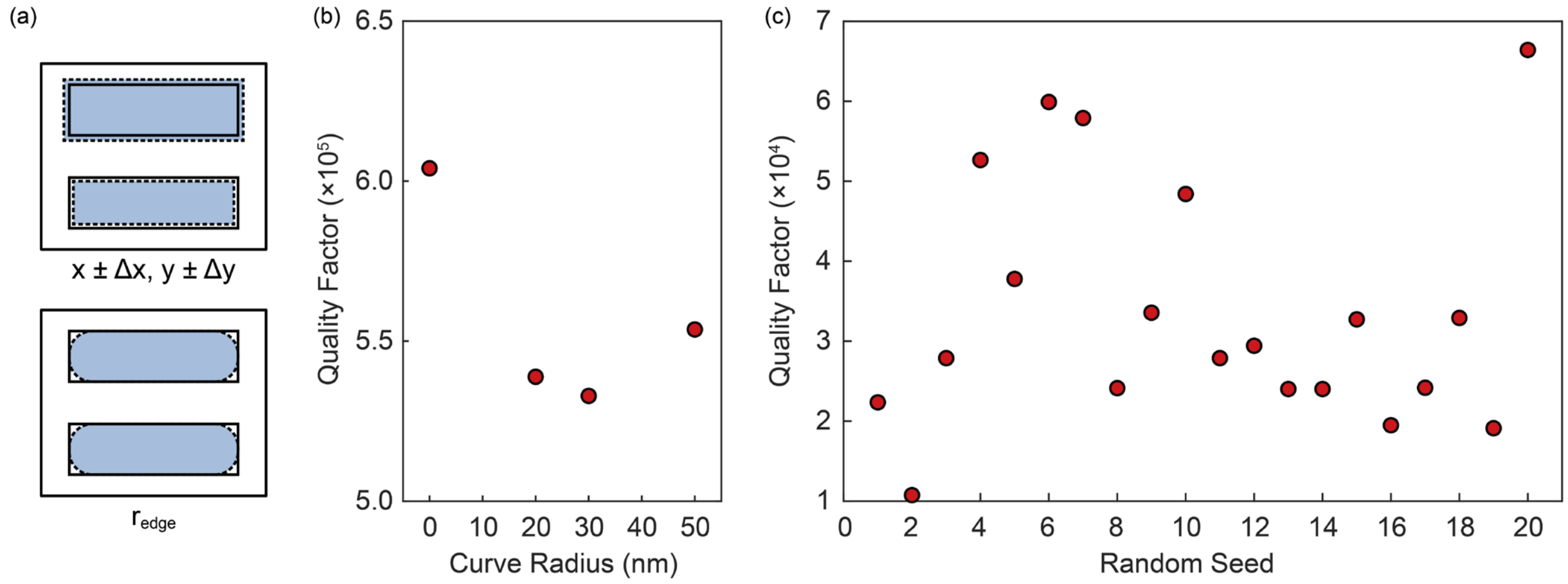

**Figure S14:** (a) Schematic illustration of two fabrication irregularities considered in eigenfrequency simulations – block-to-block dimensional mismatch (top) and rounded edges (bottom). Simulated radiative $Q$ factors of pixelated resonators with 35 nm $SiO_2$ perturbations in water are shown for (b) different edge radii and (c) multiple disorder realizations (a normal distribution of block-to-block dimensional variations with $\sigma = 3$ nm and different random seeds).

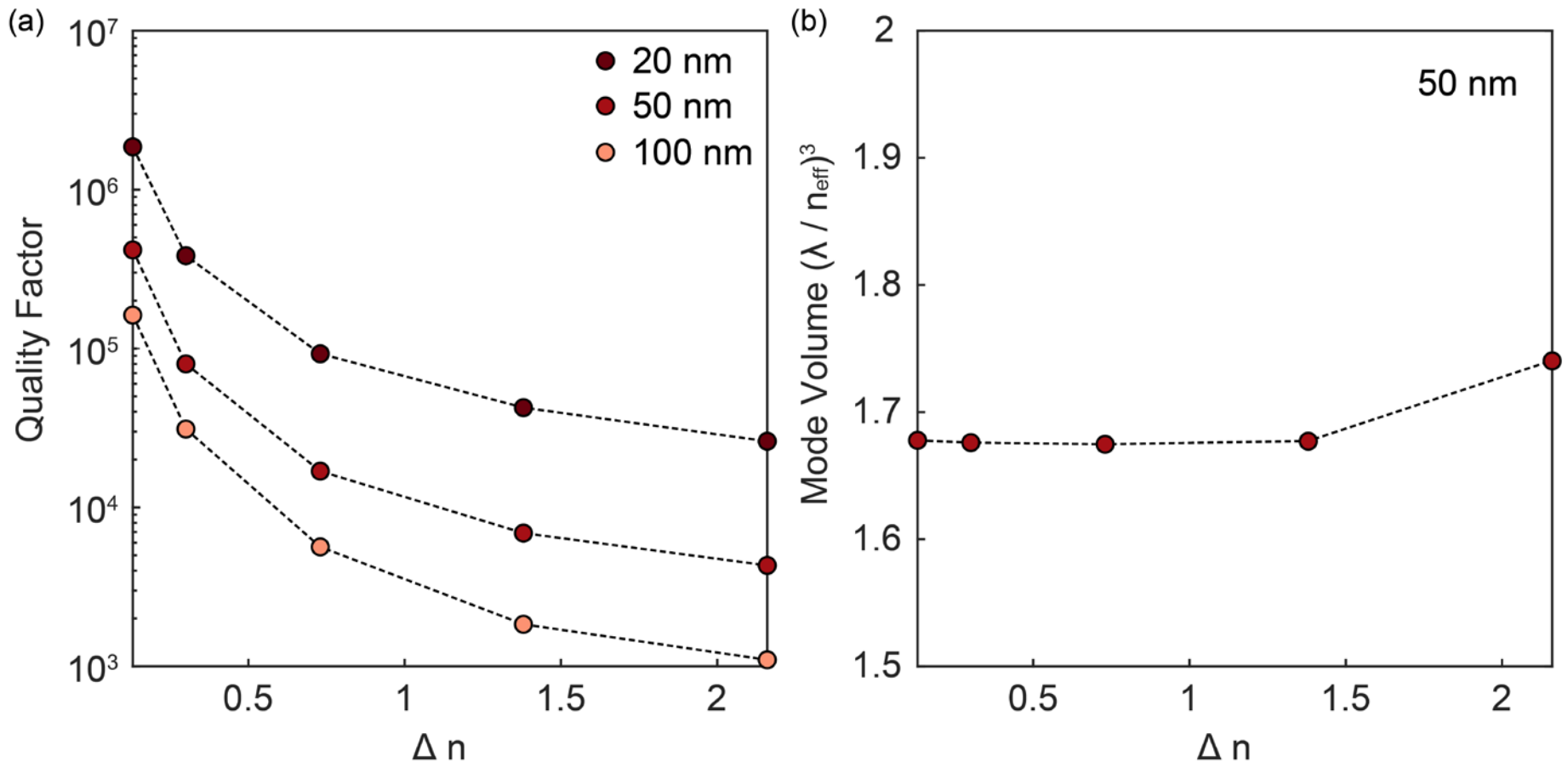

**Figure S15:** (a) Simulated $Q$ factors for pixelated resonators with 20, 50, and 100 nm height perturbations of different $\Delta n$, showing that $Q$ increases as $\Delta n \rightarrow 0$, as is the case for the infinite resonator. (b) Simulated mode volumes for pixelated resonators with 50 nm height perturbations of different $\Delta n$.